\pdfoutput=1
\documentclass[11pt]{article}

\usepackage{arxiv}

\usepackage{times}
\usepackage{longtable}
\usepackage{algorithm}
\usepackage{algorithmic}
\usepackage{nicefrac}
\usepackage{booktabs}
\usepackage{footmisc}
\usepackage[shortlabels]{enumitem}
\usepackage{multirow}
\usepackage[version=4]{mhchem}

\usepackage{array}
\usepackage{url}
\usepackage[strict]{changepage}
\usepackage{makecell}
\usepackage{titlesec}
\usepackage{setspace}
\usepackage{bm}
\usepackage{mdframed}
\usepackage{bbm}
\usepackage{threeparttable}
\usepackage{subfigure}

\usepackage[T1]{fontenc}

\usepackage{titletoc}
\usepackage{longtable}
\usepackage{pifont} 
\usepackage{wrapfig}
\usepackage{booktabs}
\usepackage[table]{xcolor}
\usepackage{makecell}

\usepackage{fancyhdr} 

\usepackage[most]{tcolorbox} 

\definecolor{darkpastelgreen}{rgb}{0.01, 0.75, 0.24}

\titleformat*{\subparagraph}{\itshape}

\newsavebox\actorsfigure

\usepackage{circledsteps}
\usepackage{fontawesome5}

\definecolor{LightGray}{RGB}{250,250,250}
\definecolor{Gray}{RGB}{240, 240, 240}

\NewDocumentCommand{\heng}
{ mO{} }{\textcolor{red}{\textsuperscript{\textit{Heng}}\textsf{\textbf{\small[#1]}}}}
\usepackage[numbers]{natbib}
\usepackage{amsfonts}
\usepackage[hidelinks]{hyperref}

\title{Bridging Brains and Machines: A Unified Frontier in Neuroscience, Artificial Intelligence, and Neuromorphic Systems}

\author{
  Sohan Shankar
  \And
  Yi Pan
  \And
  Hanqi Jiang
  \And
  Zhengliang Liu
  \And
  Mohammad R. Darbandi
  \And
  Agustin Lorenzo
  \And
  Junhao Chen
  \And
  Weihang You
  \And
  Md Mehedi Hasan
  \And
  Arif Hassan Zidan
  \And
  Eliana Gelman
  \And
  Joshua A. Konfrst
  \And
  Jillian Y. Russell
  \And
  Katelyn Fernandes
  \And
  Tianze Yang
  \And
  Yiwei Li
  \And
  Huaqin Zhao
  \And
  Afrar Jahin 
  \And
  Triparna Ganguly
  \And
  Shair Dinesha
  \And
  Yifan Zhou
   \And
  Zihao Wu
  \And
  Xinliang Li
  \And
  Lokesh Adusumilli
  \And
  Aziza Hussein
  \And
  Sagar Nookarapu
  \And
  Jixin Hou
  \And
  Kun Jiang
  \And
  Jiaxi Li
  \And
  Brenden Heinel
  \And
  XianShen Xi
  \And
  Hailey Hubbard
  \And
  Zayna Khan
  \And
  Levi Whitaker
  \And
  Ivan Cao
  \And
  Max Allgaier
  \And
  Andrew Darby
  \And
  Lin Zhao
  \And
  Lu Zhang
  \And
  Xiaoqiao Wang
  \And
  Xiang Li
  \And
  Wei Zhang
   \And
  Xiaowei Yu
  \And
  Dajiang Zhu
   \And
  Yohannes Abate
  \And
  Tianming Liu
}

\makeatletter
\renewcommand{\@maketitle}{%
  \vbox{%
    \hsize\textwidth
    \linewidth\hsize
    \vskip 0.1in
    \@toptitlebar
    \centering
    {\LARGE\sc \@title\par}
    \@bottomtitlebar
    \vskip 0.1in
    \def\And{%
      \end{tabular}\hfil\linebreak[0]\hfil%
      \begin{tabular}[t]{c}\bf\rule{\z@}{24\p@}\ignorespaces%
    }
    \def\AND{%
      \end{tabular}\hfil\linebreak[4]\hfil%
      \begin{tabular}[t]{c}\bf\rule{\z@}{24\p@}\ignorespaces%
    }
    \begin{tabular}[t]{c}\bf\rule{\z@}{24\p@}\@author\end{tabular}%
    \vskip 0.15in \@minus 0.1in%
  }
}
\makeatother

\begin{document}
\maketitle
\renewcommand{\thefootnote}{\fnsymbol{footnote}}

\footnotetext[2]{This work was undertaken as a collaborative project by students and their mentors/collaborators in the Computational Neuroscience
course (taught by Professor Tianming Liu) at The University of Georgia. Professor Tianming Liu is the current corresponding author: \href{mailto:tliu@uga.edu}{tliu@uga.edu}}
\footnotetext[3]{Latest Update: July, 2025.}


\begin{abstract}
This position and survey paper identifies the emerging convergence of neuroscience, artificial general intelligence (AGI), and neuromorphic computing toward a unified research paradigm. Using a framework grounded in brain physiology, we highlight how synaptic plasticity, sparse spike-based communication, and multimodal association provide design principles for next-generation AGI systems that potentially combine both human and machine intelligences. The review traces this evolution from early connectionist models to state-of-the-art large language models, demonstrating how key innovations like transformer attention, foundation-model pre-training, and multi-agent architectures mirror neurobiological processes like cortical mechanisms, working memory, and episodic consolidation. We then discuss emerging physical substrates capable of breaking the von Neumann bottleneck to achieve brain-scale efficiency in silicon: memristive crossbars, in-memory compute arrays, and emerging quantum and photonic devices. There are four critical challenges at this intersection: 1) integrating spiking dynamics with foundation models, 2) maintaining lifelong plasticity without catastrophic forgetting, 3) unifying language with sensorimotor learning in embodied agents, and 4) enforcing ethical safeguards in advanced neuromorphic autonomous systems. This combined perspective across neuroscience, computation, and hardware offers an integrative agenda for in each of these fields.
\end{abstract}





\newpage
\tableofcontents
\newpage
\section{Introduction}\label{Section 1}

\definecolor{quoteLine}{RGB}{0,0,0} 
\definecolor{quoteText}{RGB}{0,0,0} 
\definecolor{authorText}{RGB}{0,0,0} 
\definecolor{hidden-draw}{RGB}{106,142,189} 
\definecolor{hidden-blue}{RGB}{194,230,247} 
\definecolor{hidden-orange}{RGB}{217, 230, 252}












\subsection{Motivations}\label{subsec:1.1}
The rapid advancements in artificial intelligence (AI) and neuroscience over the past decade have reignited interest in understanding and replicating intelligence. Neuroscience has revealed intricate mechanisms of information processing in the brain, from spike-based coding to synaptic plasticity, while AI has achieved remarkable capabilities through deep neural networks (DNNs) and large language models (LLMs). At the same time, neuromorphic computing has emerged as a third pillar, aiming to build energy-efficient hardware that more closely mimics neuronal and synaptic dynamics. Yet, despite clear intersections, these three disciplines often progress in parallel rather than in concert. This survey seeks to bridge that gap by providing a unified perspective on how neuroscience, AI, and neuromorphic computing inform and inspire one another, ultimately guiding the next generation of brain-inspired intelligent systems.

Neuroscience has deepened our understanding of hierarchical memory, spiking communication, and embodied cognition. Landmark projects in connectomics and large-scale neural recordings have revealed how biological networks organize information across multiple spatial and temporal scales. Simultaneously, AI researchers have developed scaling laws for deep learning, enabling foundational models with billions of parameters that exhibit emergent behaviors. Meanwhile, neuromorphic computing platforms promise orders-of-magnitude improvements in energy efficiency by harnessing event-driven computation. Each field offers unique insights: neuroscience provides biological plausibility and learning rules; AI contributes algorithmic innovations and benchmarks; neuromorphic systems deliver hardware primitives that reflect brain-like constraints. However, existing literature often treats these areas in isolation, leaving open the question of how to form a coherent, cross-domain research agenda.

This paper aims to fill that void by surveying foundational milestones, recent advances, and conceptual mismatches across neuroscience, AI, and neuromorphic computing. Our objectives are threefold: (1) to catalog key developments in each domain and highlight historical points of cross-inspiration; (2) to analyze state-of-the-art techniques and compare their strengths and limitations; and (3) to propose an integrated roadmap that outlines open challenges and future directions for truly brain-inspired AI and next-generation neuromorphic hardware. By weaving together insights from multiple disciplines, we hope to encourage collaborative research that transcends traditional boundaries.

\subsection{Key Contributions}

The primary contributions of this survey are:
\begin{itemize}
\item A focused overview of selected classical and modern neuroscience findings (Sections 2 \& 3) and their relevance to AI concepts such as memory consolidation, attention, and plasticity.
\item An examination of recent trends in large language models, AGI benchmarks, and cognitive gaps motivating biologically inspired approaches (Section 4).
\item An introductory survey of the neuromorphic computing landscape, including key hardware architectures, spiking neural network (SNN) algorithms (e.g., surrogate-gradient training, spiking Transformers), and example applications (Section 5).
\item A conceptual discussion of points of convergence and divergence between neuroscience, AI, and neuromorphic computing, with a focus on learning rules, hierarchical memory, and embodied intelligence (Section 6).
\item A preliminary agenda outlining open research directions and challenges, from spiking-first architectures to ethical considerations for AGI (Section 8).
\end{itemize}

\subsection{Paper Roadmap}

To guide the reader through this survey, we close the introduction with a high-level roadmap of each section. In Section 2 (“Background: Milestones and Shared Principles”), we review foundational work in neuroscience (Section 2.1), the rise of AI and AGI concepts (Section 2.2), and the origins of neuromorphic engineering (Section 2.3). In particular, Section 2.4 (“Cross-Inspiration Across Fields”) highlights how early ideas in cognitive science, information theory, and analog VLSI informed modern AI, while Section 2.5 (“Convergence of Domains”) illustrates concrete points of synergy, such as shared representational formats (Section 2.5.1), learning rules (Section 2.5.2), and neuromorphic implementations of Transformer patterns (Section 2.5.3).

Building on that foundation, Section 3 (“Recent Advances in Brain Science and Neuroscience”) examines state-of-the-art connectomics and brain mapping efforts (Section 3.1), large-scale multilevel simulations (Section 3.2), and new mechanistic and developmental models of cortical organization (Section 3.3). We conclude Section 3 with “Brain-Inspired Computing Concepts” (Section 3.4), which surveys how insights from neural coding, pruning, and plasticity continue to inform today’s AI architectures.

Section 4 (“Recent Advances in Large Language Models and Developing AGI”) analyzes scaling laws and emergent capabilities in foundation models (Section 4.1), details architectural and algorithmic innovations (Section 4.2), and discusses benchmarks for evaluating intelligence (Section 4.4).We also highlight the remaining cognitive gaps in contemporary AI: embodiment (Section 4.5.1), causal reasoning (Section 4.5.2), memory (Section 4.5.3), energy efficiency (Section 4.5.4), interpretability (Section 4.5.5), and adaptability (Section 4.5.6).

In Section 5 (“Recent Advances in Neuromorphic Computing and SNNs”), we survey next-generation hardware (Section 5.1), emerging quantum-neuromorphic platforms (Section 5.2), and the evolving roles of memristors and physical neurons (Section 5.3). Section 5.4 covers algorithmic breakthroughs, from surrogate-gradient training (Section 5.4.1) to spiking Transformer architectures (Section 5.4.3) and continual on-chip learning (Section 5.4.5), while Section 5.5 showcases real-world SNN applications. We wrap up Section 5 by pinpointing ongoing engineering challenges such as device variability, software tooling, and scalability (Section 5.6).

Section 6 (“Conceptual Connections and Mismatches Across Domains”) identifies how computational frameworks (Section 6.1), learning rules (Section 6.2), and hybrid learning approaches (Section 6.3) both align and diverge when viewed through neuroscience, AI, and neuromorphic lenses. We pay special attention to memory and knowledge organization (Section 6.4) and hierarchical processing motifs that suggest new opportunities for cross-domain co-design.

Section 7 (“Implications for Cognitive Modeling, AI, and Neuroscience”) discusses broader impacts in cognitive science (Section 7.1), AGI development (Section 7.2), and neuroscience research (Section 7.3), while also surveying applications in brain-computer interfaces (Section 7.5) and neuroprosthetics. We specifically highlight how large language models can serve as proxies for brain semantics (Section 7.4) and how computational models of psychiatric disorders (Section 7.6) may emerge from these interdisciplinary efforts.

Section 8 (“Future Research Directions”) lays out an integrated agenda: biologically grounded AGI (Section 8.1), scaling neuromorphic computing (Section 8.2), core SNN-centric AGI principles (Section 8.3), and brain-computer interface synergies (Section 8.4). We conclude this section by emphasizing ethical safeguards (Section 8.5) and spelling out open challenges from spiking-first architectures to lifelong embodied intelligence (Section 8.6).

Finally, Section 9 (“Conclusion”) synthesizes our integrative agenda (Section 9.1) and offers forward-looking thoughts on how unifying neuroscience, AI, and neuromorphic computing might reshape the future of intelligent systems (Section 9.2). 
\section{Background: Milestones and Shared Principles}\label{Section 2}
\subsection{Foundations in Neuroscience and Cognitive Science}
The modern understanding of neural computing begins with Santiago Ramón y Cajal's neuronal doctrine, which established neurons as the fundamental processing units of the brain \cite{defelipe1992santiago}. Early physiological insights by Charles Sherrington, who coined the term "synapse" and characterized excitatory/inhibitory postsynaptic potentials \cite{sherringtonIntegrativeActionNervous1908}, and Otto Loewi's demonstration of chemical neurotransmission in frog-heart preparations \cite{loewiUeberHumoraleUebertragbarkeit1921}, provided the biophysical substrate for intercellular communication. Pavlov's classical conditioning laid the behavioral groundwork for associative mechanisms \cite{pavlovConditionedReflexesInvestigation1927}. A formal computational perspective emerged from McCulloch and Pitt's logical neuron model, which represented neural elements as binary threshold units wired into logical networks \cite{mccullochLogicalCalculusIdeas1943}. This foundation was built upon by Donald Hebb's seminal work, which provided the first mechanistic account of synaptic plasticity, positing that when an axon of cell A repeatedly takes part in the firing of cell B, some growth process or metabolic change increases A's efficiency in firing B. He further discussed cell assemblies, groups of neurons with strengthened mutual synapses that become co-active as a unit, and phase sequences, chains of cell assemblies activated in a temporal order  \cite{Hebb1950}. The discovery of action potentials by Hodgkin and Huxley formalized the electrochemical basis of neural signaling \cite{vandenberg2012hodgkin}, while subsequent work by Hubel and Wiesel revealed the hierarchical organization of visual processing that would later inspire convolutional architectures \cite{wurtz2009recounting}. Cognitive science contributions from Atkinson and Shiffrin's multi-store memory model \cite{atkinson1968human} and Baddeley's working memory framework \cite{baddeleyWorkingMemory1974} provided computational abstractions of memory systems that inform contemporary AI architectures. These early discoveries established critical principles: information processing through distributed neuronal networks, learning through synaptic modification, and the brain's remarkable energy efficiency despite massive parallelism. These principles remain central to both modern computational neuroscience and brain-inspired artificial intelligence systems.

\subsection{Rise of AI and AGI Concepts}
Research in artificial intelligence began in the mid-twentieth century with Alan Turing’s landmark paper bringing attention to the possibility of thinking machines \cite{turingTest}. Initial approaches to developing machines that mimic or surpass human intelligence were mainly based on logical symbol manipulation \cite{taylor2021}. This branch of AI, known as symbolic AI or good old-fashioned AI (GOFAI), solved problems by utilizing heuristics while navigating problem spaces to reach solutions faster. This was soon followed by connectionist approaches that used artificial neural networks to encode knowledge in the form of connection weights instead of explicitly defined symbolic structures \cite{smolensky1987}. Connectionist approaches are also part of a larger branch of AI known as machine learning, which is aimed at designing algorithms that use data to improve their own performance on a task over time \cite{tiwari2024}. Deep learning, which uses larger networks with many hidden layers, has been especially successful in recent decades with the advent of parallel processing and large datasets, allowing for the creation of more capable and accurate models \cite{liu2023}.

In recent years, the notion of AGI has shifted the market: whereas early visions of strong AI emphasized explicit rule-based reasoning, modern discourse increasingly characterizes general intelligence in terms of flexible, data-driven learning. Advances in neuro-symbolic integration and large-language-model research have blurred the old symbolism vs. connectionism distinction \cite{xiong2024converging}, showing that neural architectures can learn to manipulate symbols as part of language understanding. Crucially, this evolution has been driven by large-scale statistical models: contemporary “foundation models” are pre-trained on enormous multimodal corpora with self-supervised objectives and then applied across diverse tasks \cite{yuan2024power}. These models, which Fei et al. describe as possessing newfound “imagination ability,” achieve a “transformative stride towards AGI” by generalizing well beyond narrowly trained domains \cite{fei2022towards}. From this perspective, AGI is reframed not as perfect human emulation but as mastery of core cognitive skills (learning, reasoning, adaptation) via scale and architecture. For example, Tavares argues that GPT-4-like LLMs already demonstrate patterns of flexible generalization comparable to average human intelligence across modalities, attributing their capabilities to the Transformer’s inherently domain-agnostic design \cite{tavares2025pragmatic}. In line with this view, scholars note that deep learning’s “unreasonable effectiveness,” grounded in high-dimensional geometry and massive optimization, underpins emergent abilities (planning, abstraction) that were once considered AGI hallmarks \cite{raman2025navigating}. In short, each innovation – from neurosymbolic hybrids through better optimization to transformer-based pretraining – has progressively redefined AGI as the ability to generalize adaptively from data rather than rely solely on handcrafted logic \cite{fei2022towards}.

Building on the shift toward AGI, recent years have seen the rise of flagship AI models that push the boundaries of general-purpose intelligence. OpenAI’s GPT-4, Google DeepMind’s Gemini, Anthropic’s Claude, and Meta’s LLaMA are among the most prominent examples, each heralded as a major advance in versatile AI capabilities. Unlike earlier systems tailored to single tasks, these cutting-edge models are trained on unprecedentedly broad datasets, equipping them to perform an array of functions—from complex reasoning and creative writing to coding and data analysis—within a single unified architecture. Often described as “foundation models,” they are designed for adaptability across domains and can be fine-tuned to a wide range of downstream tasks, making them prime examples of the new wave of general-purpose AI shaping the field.

Collectively, these models illustrate a broader conceptual shift from narrow, specialized AI toward more general and adaptable systems. All of them leverage massive neural networks to capture patterns across diverse types of data, and some are explicitly built to integrate multiple modalities (for instance, Gemini was designed from the ground up to be multimodal, capable of handling text, code, audio, images, and video in one model \cite{Google2023Gemini}. This means a single AI system can now generalize across tasks and formats, mirroring a flexibility once seen only in human cognition. Moreover, progress is not limited to any one organization’s approach: while GPT-4 and Claude have set new benchmarks in conversational and reasoning ability, Meta’s LLaMA opened access to large-scale models through an open-source release, and Google’s Gemini is charting advances in real-time reasoning and tool use – all reflecting a convergent trend toward AI that is more universal in scope and application. Each successive model generation extends core capacities (e.g., greater context “memory,” creativity, or safety alignment), underscoring how scaling up architectures and training can yield emergent capabilities that bring AI closer to generalized problem-solving.

Significantly, the capabilities of these flagship models are beginning to encroach on human-like competence in various domains. In some evaluations, current top models achieve near-human levels of comprehension and fluency on complex tasks, placing them at the frontier of general intelligence \cite{Anthropic2024Claude3}. This marks a pivotal moment in AI: the gap between specialized narrow AI and more human-like general intelligence is visibly narrowing. The rise of such versatile systems underscores a paradigm shift in how we conceive AI – not as isolated expert programs, but as general problem-solvers with the capacity to adapt and learn broadly. While challenges and limitations remain (indeed, even the most advanced models can falter outside their training distribution or in high-stakes settings), their impressive breadth of competence fuels optimism that increasingly unified AI systems can be developed. In essence, today’s GPT-4, Gemini, Claude, LLaMA and their peers serve as both milestones and catalysts – concrete milestones of progress toward more adaptive, general AI, and catalysts prompting deeper inquiry into what it will take to fully bridge the gap between artificial and natural intelligence in the coming years \cite{WikipediaFoundationModel}. 

\subsection{Origins of Neuromorphic Computing and SNNs}
Neuromorphic engineering emerged conceptually in the 1980s, largely driven by the pioneering work of Carver Mead at the California Institute of Technology. Mead explored the potential of analog circuits designed to function analogously to biological neurons \cite{mead1989analog}. In 1989, building upon this foundation, he introduced the term “neuromorphic” to describe electronic systems explicitly modeled on the structure and function of biological nervous systems \cite{mead1990neuromorphic}. A key early success stemming from this research was the silicon retina, developed by Mead's student, Misha Mahowald, which functionally mimicked aspects of biological visual processing \cite{mahowald1988silicon, mahowald1992thesis}. Subsequent research efforts focused on constructing silicon emulations of other sensory systems, such as the inner ear, and developing various models of spiking neurons. A defining characteristic of these early neuromorphic designs was their reliance on the intrinsic physical properties and electrical behavior of silicon devices, particularly transistors operating in the subthreshold regime, rather than implementing complex digital logic to simulate neural functions. This approach inherently favored massively parallel operation, continuous-valued analog signals, and event-driven processing, contrasting sharply with the sequential, discrete, and clock-driven nature of conventional von Neumann computers \cite{indiveri2011neuromorphic}.

\subsubsection{The Genesis of Neuromorphic Engineering: Carver Mead and Analog VLSI}
The conceptual origins of neuromorphic engineering are inextricably linked to Carver Mead's research trajectory in the 1980s. Mead recognized a fundamental disparity: biological nervous systems, even in relatively simple organisms, demonstrated computational capabilities for tasks like perception and motor control that far surpassed those of contemporary digital computers, despite the latter's advantages in raw processing speed \cite{mead1989analog, mead1990neuromorphic}. This observation prompted a shift in perspective, moving away from purely digital abstractions towards understanding and harnessing the computational principles inherent in biological systems.

Central to Mead's vision was the use of Analog Very Large Scale Integration (VLSI) technology, as both a substrate for implementing digital logic and as a computational medium in its own right \cite{mead1989analog}. He particularly focused on the subthreshold, or weak inversion, operating regime of Metal-Oxide-Semiconductor (MOS) transistors. In this regime, the drain current exhibits an exponential dependence on the gate voltage, a characteristic highly similar to the relationship between ion flow and membrane potential difference governed by the Nernst equation in biological ion channels \cite{mead1989analog, indiveri2011neuromorphic}. This was a radical departure from conventional digital VLSI design, which primarily utilized transistors as binary switches, effectively abstracting away their underlying analog physics \cite{mead1989analog}. Mead proposed using this intrinsic analog behavior directly for computation, enabling elegant and power-efficient implementations of functions like exponentiation, logarithms, and multiplication directly within the physics of the silicon \cite{mead1989analog, mead1990neuromorphic}. His seminal 1989 book, Analog VLSI and Neural Systems, provided a foundational exposition of these principles, detailing circuits that emulated sensory processing, famously including "silicon retinas" and "silicon cochleas" \cite{mead1989analog}. This work suggested that computation could emerge directly from the physical properties of the hardware substrate, offering potential efficiency gains.

Around 1989-1990, Mead formalized this approach by coining the term "neuromorphic" \cite{mead1990neuromorphic, indiveri2011neuromorphic}. This term specifically denoted electronic systems whose architecture and operational principles were directly inspired by, and sought to emulate, those found in biological nervous systems \cite{mead1990neuromorphic}. It captured the core philosophy of building hardware that mirrored the brain's strategies, establishing the key tenets of the budding field: leveraging analog computation derived from device physics, embracing the massive parallelism inherent in neural structures, and adopting event-driven processing where computation and communication occur asynchronously in response to significant events (spikes), rather than synchronizing through a global clock.

\subsubsection{Early Hardware Manifestations: Silicon Neurons and Bio-Inspired Sensors}
Building upon Mead's foundational principles, the 1990s saw the emergence of tangible hardware implementations, most notably physical "silicon neurons" (SiNs) and bio-inspired sensory systems. SiNs were typically realized as mixed-signal (analog/digital) VLSI circuits designed to replicate the electrophysiological dynamics of their biological counterparts \cite{indiveri2011neuromorphic, mahowald1991silicon}. Among the earliest and most common models implemented were the Integrate-and-Fire (I \& F) and its variant, the Leaky Integrate-and-Fire (LIF) neuron \cite{indiveri2011neuromorphic, burkitt2006review}. These circuits typically used a capacitor to represent the neuronal membrane, integrating input currents (analogous to synaptic inputs) over time. When the voltage across the capacitor (membrane potential) reached a predefined threshold, the circuit generated a digital pulse (an action potential or spike), followed by a reset of the membrane potential and often a brief refractory period during which spiking was suppressed \cite{indiveri2011neuromorphic, burkitt2006review}. While computationally simple and relatively easy to implement in silicon, these early models had limitations, such as not capturing the precise shape of biological spikes and sometimes having fixed time constants determined by the fabrication process \cite{indiveri2011neuromorphic}. Still, they provided crucial building blocks for constructing larger neuromorphic systems.

A landmark achievement of this era was the silicon retina, developed primarily by Misha Mahowald \cite{mahowald1988silicon, mahowald1992thesis}. This device transcended simple photodetection; it was an analog VLSI chip that performed computations analogous to those occurring in the outer plexiform layer of the vertebrate retina \cite{mahowald1988silicon, boahen1992contrast}. The circuit incorporated photoreceptor elements that computed a logarithmic response to incident light intensity, effectively compressing a wide dynamic range. It utilized resistive networks to compute spatially smoothed versions of the photoreceptor outputs, mimicking the function of horizontal cells, and subsequently computed the contrast by taking the difference between the local photoreceptor signal and the smoothed signal, analogous to bipolar cell responses \cite{mahowald1988silicon, boahen1992contrast}. This system demonstrated the power of analog VLSI to implement complex, biologically inspired processing directly on-chip, performing tasks like dynamic range adaptation and edge enhancement efficiently in hardware \cite{mahowald1988silicon, mahowald1992thesis}. The success of the silicon retina spurred the development of other neuromorphic sensors, such as the silicon cochlea, which modeled the hydrodynamics and frequency decomposition performed by the inner ear \cite{mead1989analog}.

Simultaneously, researchers began tackling the challenge of implementing learning and adaptation in hardware. Inspired by Donald Hebb's postulate that correlated activity strengthens synaptic connections ("neurons that fire together, wire together") \cite{hebb1949organization}, efforts focused on realizing synaptic plasticity mechanisms in silicon. Spike-Timing-Dependent Plasticity (STDP) emerged as a particularly influential biological phenomenon and computational model \cite{gerstner1996neuronal, bi1998synaptic}. In STDP, the change in synaptic strength (weight) depends critically on the precise relative timing of pre-synaptic and post-synaptic spikes within a narrow temporal window. If a pre-synaptic spike consistently precedes a post-synaptic spike, the synapse tends to strengthen (Long-Term Potentiation, LTP); if it consistently follows, the synapse tends to weaken (Long-Term Depression, LTD). Early neuromorphic implementations explored various analog and mixed-signal circuit designs to capture this temporal dependency and implement these local, unsupervised learning rules directly in hardware \cite{bofill2004vlsi, indiveri2006vlsi}. This exploration of hardware-based local learning rules represented a significant divergence from traditional machine learning approaches often reliant on global optimization algorithms like backpropagation. It aligned naturally with the distributed architecture of neuromorphic systems, where computation and memory (synaptic state) are co-located, potentially enabling continuous, online adaptation without requiring extensive data shuffling or powerful central controllers.

As these neuromorphic chips incorporating sensors, neurons, and plasticity mechanisms grew in complexity and scale, efficient inter-chip communication became paramount. The sparse and asynchronous nature of spiking activity rendered traditional synchronous bus communication inefficient. The Address Event Representation (AER) communication protocol was developed, with pioneering contributions from Sivilotti and Mahowald to help resolve this issue \cite{mahowald1992thesis, boahen2000point, sivilotti1991thesis}. AER utilizes a shared digital bus to transmit the unique address or identifier of a neuron whenever it generates a spike \cite{boahen2000point}. Access to the shared bus is managed by an asynchronous handshake protocol (typically involving request /REQ and acknowledge /ACK signals), allowing events to be transmitted as soon as they occur \cite{boahen2000point}. This time-division multiplexing approach leverages the significant speed advantage of electronic signaling over biological spiking frequencies, enabling a large number of virtual neuronal connections to be routed over a small number of physical wires (scaling logarithmically, O(logN), rather than linearly, O(N)) \cite{boahen2000point, mahowald1992thesis}. The concurrent development of bio-inspired sensors providing relevant, pre-processed inputs, silicon neurons performing the core computation, local learning rules enabling adaptation, and AER facilitating efficient communication illustrates a co-evolutionary process. Progress in neuromorphic engineering required, and continues to require, synergistic advancements across sensing, computation, learning, and communication, all tailored to the event-based, parallel, and physically embodied nature of the paradigm.

\subsubsection{Architectural Advantages Over Von Neumann Architectures}
A primary motivation for developing neuromorphic computing architectures stems from the inherent limitations of the prevailing von Neumann architecture used in conventional computers. The von Neumann model separates the CPU from the main memory (RAM), requiring data and instructions to be constantly transferred back and forth across a connecting bus \cite{backus1978can}. This physical separation and the sequential nature of data transfer create the "von Neumann bottleneck," which limits computational throughput and consumes substantial energy, particularly for data-intensive applications frequently encountered in areas like artificial intelligence and real-time signal processing.

Neuromorphic architectures offer a fundamentally different approach designed to mitigate this bottleneck. Inspired by the structure of the brain, they typically feature:
\begin{enumerate}
\item \textbf{Co-location of Memory and Processing:} Artificial neurons (processing units) and their associated synaptic weights (memory elements) are tightly integrated, often physically adjacent on the silicon die \cite{indiveri2011neuromorphic}. This drastically reduces the distance and time required for data movement between memory and computation.
\item \textbf{Massive Parallelism:} Systems are composed of a large number of relatively simple processing units (neurons and synapses) operating concurrently, mirroring the parallel organization of the brain \cite{mead1990neuromorphic}.
\item \textbf{Event-Driven and Sparse Communication:} Information is typically encoded in asynchronous events (spikes). Communication protocols like AER ensure that data is transmitted across the system only when a neuron becomes active, capitalizing on the often sparse nature of neural activity \cite{boahen2000point}. This contrasts sharply with the continuous, clock-driven data transfers in synchronous digital systems.
\end{enumerate}
These architectural distinctions yield significant potential advantages:
\begin{itemize}
\item \textbf{Energy Efficiency:} By minimizing data movement and activating circuits only when processing relevant events, neuromorphic systems can achieve substantially lower power consumption compared to simulating equivalent neural networks on von Neumann machines, potentially by orders of magnitude for suitable tasks \cite{mead1990neuromorphic,indiveri2011neuromorphic}. Early estimates suggested potential efficiencies reaching the order of 10-15 Joules per synaptic operation might be achievable in silicon \cite{mead1990neuromorphic}.
\item \textbf{Speed and Real-Time Processing:} The inherent parallelism allows neuromorphic hardware to perform computations, particularly those involving spatio-temporal pattern recognition or complex simulations, much faster than sequential processors, often enabling real-time operation independent of network size \cite{indiveri2011neuromorphic}.
\item \textbf{Robustness and Fault Tolerance:} The distributed nature of the computation and the use of analog components can lend neuromorphic systems a degree of resilience to noise, component variations, and even outright failures of individual units, mirroring the robustness observed in biological neural systems \cite{mead1990neuromorphic}.
\end{itemize}

However, the advantages extend beyond simply improving speed and power efficiency for existing computational tasks. The neuromorphic approach represents a distinct computational paradigm potentially better suited for problems that interface directly with the physical world, involve processing noisy and incomplete sensory data, require continuous adaptation and learning, or demand real-time responsiveness. While von Neumann architectures excel at high-precision, symbolic, and sequential computation, the parallelism, analog processing, event-based nature, and capacity for local learning in neuromorphic systems align more naturally with the demands of tasks like robotic control, real-time sensor fusion, anomaly detection in complex data streams, and adaptive signal processing. Therefore, the "advantage" is not merely quantitative but also qualitative, representing a potentially superior architectural match for a specific, important class of computational problems where traditional methods struggle or are inefficient.

\subsubsection{Evolution Towards Large-Scale Neuromorphic Projects}
Following the foundational work and early chip demonstrations of the 1980s and 1990s, the 2000s and 2010s marked a significant transition towards the development of large-scale neuromorphic systems. These projects aimed to move beyond small-scale proofs-of-concept to build hardware platforms capable of simulating substantial portions of neural tissue or tackling complex, real-world computational challenges, leveraging the principles of brain-inspired computation at unprecedented scales.

This era saw the emergence of several influential large-scale projects, often supported by major research initiatives or industrial investment. These platforms, while sharing the common goal of brain-inspired computation, explored diverse architectural philosophies and technological approaches, reflecting the ongoing search for optimal design strategies. Key examples include:
\begin{itemize}
    \item \textbf{Neurogrid (Stanford University, ~2014):}
 This system utilized a mixed-signal approach, employing analog subthreshold circuits within specialized "Neurocore" chips for neuron and synapse emulation to achieve high biological realism and energy efficiency, while using digital circuits for spike communication (AER). Neurogrid was designed to simulate up to 1 million neurons and billions of synapses in real-time, targeting computationally demanding neuroscience modeling \cite{benjamin2014neurogrid}.
\item \textbf{TrueNorth (IBM, DARPA SyNAPSE Program, ~2014):} Representing a fully digital approach, TrueNorth integrated 1 million programmable spiking neurons and 256 million configurable synapses on a single chip. It emphasized extreme low power consumption (tens of milliwatts) and scalability through seamless tiling of multiple chips. Its architecture was optimized for event-driven processing and aimed at cognitive computing tasks \cite{merolla2014million}.
\item \textbf{SpiNNaker (Spiking Neural Network Architecture, University of Manchester, EU Human Brain Project, ~2011-ongoing):} This project adopted a massively parallel digital approach, using a large number of custom multi-core processors (based on ARM architecture) specifically optimized for simulating SNNs. SpiNNaker offers significant flexibility and programmability, designed to simulate very large networks (approaching billions of neurons across large installations) in biological real-time, primarily serving as a tool for computational neuroscience research \cite{furber2014spinnaker}.
\item \textbf{BrainScaleS (University of Heidelberg, EU Human Brain Project, ~2011-ongoing):} In contrast to SpiNNaker, BrainScaleS employs physical (analog) models of neurons and synapses implemented directly in wafer-scale VLSI technology. A key feature is its accelerated operation, running simulations up to 10,000 times faster than biological real-time. This acceleration is particularly valuable for studying slow processes like synaptic plasticity and long-term learning \cite{schemmel2010wafer}.
\item \textbf{Loihi (Intel Labs, ~2017-ongoing):} Intel's Loihi represents a digital, asynchronous neuromorphic research processor. It features highly configurable neuron models, integrated on-chip learning capabilities based on programmable synaptic plasticity rules, and efficient hierarchical mesh networks for spike communication. Loihi chips can be scaled into larger systems (e.g., Pohoiki Beach, Kapoho Point) and are aimed at exploring the potential of neuromorphic computing for AI workloads requiring low power and real-time learning \cite{davies2018loihi}.
\end{itemize}
The development of these large-scale platforms underscored the significant progress in neuromorphic engineering but also brought new challenges to the forefront. Creating effective software tools, programming models, and algorithms tailored to these non-von Neumann architectures became critical bottlenecks \cite{davies2018loihi}. Effectively harnessing the potential of on-chip plasticity for real-world learning tasks also remains an active area of research.
The diversity among these major projects highlights that the field did not converge on a single "best" architecture during this scaling phase. Instead, these efforts represent distinct exploration paths, each prioritizing different aspects of brain-inspired computation – analog versus digital implementations, biological realism versus computational abstraction, real-time versus accelerated simulation, flexibility versus specialization, and the integration of on-chip learning. This architectural divergence signifies an ongoing, vibrant search within the research community to identify the most effective and scalable strategies for building powerful and efficient neuromorphic computing systems.

\subsection{Cross-Inspiration Across Fields}
The evolution of fields concerned with understanding and replicating intelligence, which spans neuroscience, cognitive psychology, artificial intelligence (AI), information theory, and engineering, has been characterized by a profound and often reciprocal exchange of ideas. Early theoretical neurons such as the McCulloch-Pitts unit laid the logical foundation for both cognitive neuroscience and machine computation \cite{mcculloch1943}. Hebb's postulate that "cells that fire together wire together" not only reshaped the theories of synaptic plasticity but also became the algorithmic core of Hebbian learning in artificial networks and the associative-memory dynamics of Hopfield networks \cite{Hebb1950,hopfieldNeuralNetworksPhysical1982}. Information-theoretic work by Shannon, Attneave, and Barlow framed neural coding as an efficiency problem, anticipating today's sparse-coding and predictive-coding models \cite{shannonMathematicalTheoryCommunication,attneaveInformationalAspectsVisual1954,barlowPossiblePrinciplesUnderlying}. Rosenblatt's perceptron translated these biological insights into a trainable classification machine, inaugurating modern connections \cite{rosenblattPerceptronProbabilisticModel1958}. Cognitive-psychology constructs such as distributed memory systems and attentional gating later inspired architectural motifs in AI, while Bayesian statistical inference provided a normative lens on the perception and sensorimotor control, the so-called "Bayesian-brain" hypothesis \cite{knillBayesianBrainRole2004}. Periods of stagnation in digital AI (e.g., the first "AI winter" following critiques by \cite{minskyPerceptronsIntroductionComputational}) motivated researchers to revisit biological plausibility and energy efficiency. At the same time, signal-processing advances from electrical engineering allowed large-scale analyses of neural datasets (e.g., EEG, multi-unit spike sorting). This cross-influence crystallized in Carver Mead's neuromorphic engineering, where VLSI circuits such as the "silicon retina" and "silicon cochlea" exploited transistor physics to emulate neural dynamics with extreme power efficiency \cite{meadAnalogVLSIImplementation1989}. Kohonen's self-organising maps, modelled on cortical sensory topographies, further exemplify this interrelatedness \cite{kohonenSelforganizedFormationTopologically1982}. Thus, the continued evolution in the fields of neuroscience, cognitive science, AI, information theory, and engineering is best understood as a synergistic loop, with breakthroughs in one field allowing further achievements in the others. 

\subsection{Convergence of Domains}\label{subsec:convergence}
This section will discuss four pivotal areas where these domains intersect and mutually reinforce progress: shared computational and representational principles, learning rules and synaptic plasticity, neuromorphic implementations of AI patterns, and the influence of connectomics and wiring economy.

\subsubsection{Shared Computational and Representational Principles}\label{subsubsec:shared_principles}
A significant point of convergence lies in the observation that computational strategies and representational formats employed by modern large-scale AI models show remarkable parallels with those observed in biological neural systems, particularly in the context of language processing and attention.

Modern LLMs, especially those based on the Transformer architecture, develop deep, hierarchical representations of language \cite{vaswani2017attention}. These internal representations exhibit striking similarities to the patterns of neural activity observed in the human language network during the processing of syntactic and semantic information \cite{schrimpfNeuralArchitectureLanguage2021,hosseiniArtificialNeuralNetwork2024}. Studies where LLMs are exposed to the same linguistic stimuli (e.g., sentences, narratives) as human participants have revealed that the activation patterns within these models can account for a substantial portion of the variance in recorded neural responses \cite{schrimpfNeuralArchitectureLanguage2021}. This suggests that the statistical learning processes underlying LLMs capture salient aspects of how the brain encodes and processes language. Further supporting this link, LLMs have been successfully employed as computational proxies to model human semantic representation in the visual domain. By using LLM-derived semantic features (extracted via Visual Question Answering) to predict functional magnetic resonance imaging (fMRI) responses to natural images, researchers have validated the approach against known neural patterns and uncovered hierarchical semantic organizations across cortical regions \cite{liuTalkingBrainUsing}. The alignment between LLM representations and brain activity appears particularly strong during the early phases of model training, coinciding with the model mastering formal linguistic tasks like grammatical judgments. However, as models continue training and surpass human-level proficiency on the next-word prediction task, this alignment with both neural activity and human behavioral patterns (like reading times) tends to diminish \cite{ohWhyDoesSurprisal2022}. This divergence suggests that the optimization objectives currently used for training LLMs may push models towards statistical regularities or processing strategies that differ from those underlying human language comprehension, particularly once a certain level of predictive capability is exceeded. It implies that while sequence prediction captures fundamental aspects of linguistic structure, human language processing likely involves additional constraints, objectives, or mechanisms not fully encompassed by this single training paradigm.

The self-attention mechanism, a cornerstone of the Transformer architecture \cite{vaswani2017attention}, provides another compelling example of neuro-AI convergence, having been directly inspired by neuroscientific theories of attention and neural synchrony. However, the biological plausibility of the standard dot-product softmax attention, because it requires exhaustive query–key comparisons and a global normalization, has been flagged as challenging for neuron-only circuits \cite{kozachkovBuildingTransformersNeurons2023}. Recent work, however, shows that the same computation can be realised via astrocyte-modulated tripartite synapses, effectively a biologically viable soft-max mechanism \cite{kozachkovBuildingTransformersNeurons2023}. Another implementation, the "match-and-control" principle, proposes that transformer-like attention can be implemented via short-term Hebbian plasticity occurring at the dendritic spine level \cite{ellwoodShorttermHebbianLearning2024}. In this model, the similarity between a query (represented by somatic back-propagating action potentials) and multiple keys (represented by presynaptic spike trains at apical dendrites) is computed locally via calcium dynamics mediated by NMDA receptors. Spines detecting a strong match undergo transient potentiation, allowing the corresponding key's associated value (also encoded in the presynaptic spike train) to control the neuron's output \cite{ellwoodShorttermHebbianLearning2024}. Other approaches, like the SPARKS model, adapt self-attention using Hebbian learning principles (specifically, STDP-like mechanisms with eligibility traces) to directly process the timing of single spikes and neuronal firing sequences \cite{skatchkovskyBiologicallyInspiredAttention2024}. This evolution from high-level conceptual inspiration (as in the original Transformer) towards detailed, mechanistic models grounded in synaptic biophysics signifies a deepening engagement with neural implementation details, aiming to bridge the gap between functional equivalence and biological feasibility.

Beyond individual mechanisms, the architectural organization of the brain, particularly the cerebral cortex, is increasingly informing AI model design. Recognizing that biological computation is often localized and hierarchical, researchers are incorporating these structural priors. For example, some self-attention variants inject cortical-style locality by mapping each attention head onto segregated microcircuit motifs—e.g., using cortico-thalamo-cortical loops to enforce local interactions \citep{granierMultiheadSelfattentionCorticothalamic2025,mullerTransformersCorticalWaves2024}. Similarly, brain-inspired convolutional architectures, such as those using Lp-Convolution, have been shown to benefit large kernel sizes and align better with processing in the visual cortex \cite{kwon2025brain}. The complex folding patterns of the cortex (gyrification) also hint at underlying organizational principles related to connectivity and processing efficiency \cite{zhang2017mechanisms}. Furthermore, the hierarchical nature of representations learned by LLMs resonates with the hierarchical processing streams observed in sensory cortices and language networks, as evidenced by studies mapping semantic hierarchies onto cortical activity \cite{huthNaturalSpeechReveals2016}. This focus on network structure and topology, moving beyond just the properties of individual neurons, reflects a growing appreciation that the organization of computation is as critical as the computational units themselves for achieving brain-like efficiency and capability.

Efforts to bridge the gap between the powerful representational capabilities of ANNs like Transformers and the biological plausibility and energy efficiency of Spiking Neural Networks (SNNs) have led to hybrid architectures. "Brain Transformers" represent a notable example, integrating SNN layers, which operate using discrete, temporally precise spikes similar to biological neurons with conventional attention mechanisms. Developing such models necessitates designing SNN-compatible analogues of core Transformer components, including approximations for matrix multiplication (SNNMatmul), the softmax function (SNNSoftmax), and non-linear activation functions like SiLU (SNNSiLU), often leveraging the temporal dynamics and event-driven nature of SNNs. A 3-billion parameter implementation, BrainTransformers-3B-Chat, has demonstrated competitive performance on standard LLM benchmarks such as MMLU, BBH, ARC-C, and GSM8K, suggesting the viability of this hybrid approach \cite{tang2024braintransformers}.

A more radical approach to convergence involves utilizing biological matter itself as a computational substrate. Organoid Intelligence (OI) explores the potential of using lab-grown, three-dimensional brain organoids derived from human stem cells as platforms for computation and learning \cite{smirnovaOrganoidIntelligenceOI2023}. By applying patterned electrical, chemical, or optical stimulation and analyzing the resulting electrophysiological data with machine-learning algorithms, researchers can probe organoid plasticity, map stimulus–response relationships, and build predictive models for drug efficacy and neurotoxicity screening. Beyond biomedical applications, OI platforms are being explored for sustainable bioengineering challenges such as carbon capture and bioremediation, taking advantage of organoids' innate metabolic pathways \cite{wadanOrganoidIntelligenceBiocomputing2025}. OI represents a fundamental shift from \textit{in silico} simulation or emulation towards \textit{in vitro} biological computation. While this could provide unprecedented insights into the mechanisms of biological learning, development, and potentially cognition, it concurrently raises profound ethical questions regarding consciousness, moral status, consent, and data security that are distinct from those associated with traditional silicon-based AI \cite{hyunEthicalIssuesRelated2020}.

\subsubsection{Learning Rules and Synaptic Plasticity}
For artificial neural networks (ANNs), global backpropagation is the default learning rule, while biological brains rely on local plasticity laws. Incorporating biologically-inspired spike-timing-dependent plasticity (STDP)—where relative timing of pre- and post-synaptic spikes dictates synaptic weight changes—and its higher-order, reward-modulated variants have been shown to support online continual learning in SNNs while mitigating catastrophic forgetting \cite{khoeeMetaLearningSpikingNeural2023}. The development of neuromorphic hardware provides critical platforms for investigating these rules. For example, Intel’s Loihi 2 neuromorphic research chip demonstrates on-chip STDP with nanosecond-scale weight updates, providing a live testbed for brain-inspired learning directly in silicon \cite{davies2023loihi2}. Recognizing the respective strengths of ANNs and SNNs, hybrid training methodologies are gaining traction. These often involve pre-training deep ANN architectures using standard techniques and then fine-tuning them as SNNs using surrogate gradient methods, which approximate the non-differentiable spiking mechanism to enable backpropagation-like learning. This approach aims to combine the scalability and performance achieved through deep learning advancements with the event-driven efficiency and temporal processing capabilities of SNNs \cite{yueHybridSpikingNeural2023}.

\subsubsection{Neuromorphic Implementations of AI Patterns}
	A significant thrust in the convergence involves mapping successful large-scale AI architectures, developed primarily for conventional hardware (CPUs/GPUs), onto specialized neuromorphic computing substrates. The primary motivation is to harness the potential for dramatic improvements in energy efficiency offered by the event-driven, asynchronous, and massively parallel nature of neuromorphic hardware. Notable progress has been made in translating complex models like components of the GPT architecture into SNN equivalents. For instance, SpikeGPT, an SNN implementation of a GPT-style decoder, demonstrated remarkable energy efficiency—reportedly up to 22 times higher than its GPU-based counterpart—on tasks like next-token prediction, highlighting the potential energy savings for large model inference \cite{zhuSpikeGPTGenerativePretrained2024}. Other SNN adaptations of Transformer-based models, such as Spikformer \cite{zhouSpikformerWhenSpiking2022} and SpikingBERT \cite{balSpikingBERTDistillingBERT2024}, further illustrate this trend. To facilitate this transition, standardized software flows are being developed. Tools enabling the conversion of models defined in popular machine learning frameworks (like PyTorch, via the Open Neural Network Exchange - ONNX format) into SNNs deployable on various neuromorphic platforms (e.g., Intel's Loihi, SpiNNaker, BrainScaleS) are becoming increasingly crucial \cite{pedersenNeuromorphicIntermediateRepresentation2024}. These tools help bridge the gap between the AI development ecosystem and the specialized neuromorphic hardware landscape, although challenges remain in preserving model accuracy and performance post-conversion and adapting models to diverse neuron dynamics \cite{hu2024toward}.

\subsubsection{Connectomics and Wiring Economics}

    Advances in neuroscience, especially whole-brain connectomics, are generating principles that feed back into AI research. One such principle is "wiring economy," the observation that biological neural networks appear optimized to minimize wiring costs (e.g., total length of neuronal processes, metabolic energy). Incorporating principles of wiring economy into ANN design, for instance, through regularization techniques that penalize connection length or promote sparsity, has been shown to improve model efficiency, robustness, and generalization capabilities \cite{zhangBraininspiredWiringEconomics2025}. Constraints related to physical embedding and minimizing wiring length also influence phenomena like cortical folding \cite{zhang2017mechanisms}. Biologically inspired graph pruning strategies, which selectively remove connections or neurons based on criteria derived from neuroscience, are reshaping our understanding of how sparse yet powerful representations can be achieved efficiently in both silicon hardware and biological tissue. Neuroscientific data can inform AI architectures, and AI tools can help analyze neuroscientific data, creating a feedback loop that accelerates our understanding of intelligence. 
\section{Recent Advances in Brain Science and Neuroscience}\label{Section 3}

\subsection{Connectomics \& Brain Mapping}
Understanding the brain’s complex wiring architecture is a fundamental goal of modern neuroscience, and recent advances in connectomics are beginning to make whole-brain cellular-resolution circuit mapping a reality~\cite{peng2021morphological}. Recent breakthroughs in neuron morphology reconstruction and data integration have ushered in a new era of brain mapping, laying the foundation for both computational modeling and biologically grounded AGI.

One of the landmark efforts in this space is the BRAIN Initiative Cell Census Network (BICCN), which combines transcriptomics, epigenomics, spatial transcriptomics, and full single-neuron reconstructions to build a multimodal atlas of the brain’s cellular organization. In particular, the study by Peng et al. reconstructed 1,741 neurons across the mouse brain and classified them into 11 major projection neuron types~\cite{peng2021morphological}, revealing principles of axonal projection diversity and morphological variance that transcend transcriptomic clustering alone. This dataset highlights the intricate and non-linear relationships between gene expression and structural connectivity, and underscores the need for cross-modality data to fully characterize cell types~\cite{biccn2021multimodal}.

Complementary to this effort, the multimodal MOp atlas further integrated transcriptomics, DNA methylation, chromatin accessibility, morphology, and physiology to build a hierarchical taxonomy of neuronal types across species (mouse, marmoset, and human). This work not only provides a conserved classification framework but also links molecular identity with electrophysiological and connectional properties~\cite{biccn2021multimodal}. Notably, axonal projection diversity was found to be only partially explained by transcriptomic identity, indicating additional layers of regulation that shape neural circuits and necessitate whole-cell reconstructions for ground-truth validation.

To facilitate open access and dynamic querying of morphological data, platforms such as NeuroXiv have been developed. NeuroXiv~\cite{jiang2025neuroxiv} currently hosts over 175,000 standardized neuron reconstructions, all mapped to the Allen Common Coordinate Framework (CCFv3), and integrates AI-powered data mining for real-time exploration and hypothesis generation. Its engine allows users to perform morphometric analyses, identify projection motifs, and study dendritic and axonal arborization across brain regions and cell types.

Further scaling efforts have culminated in datasets like the SEU-A1876 from over 500 mouse brains, offering peta-voxel resolution imaging and full morphological reconstructions for 1,876 neurons, alongside millions of putative synaptic boutons. This dataset supports multi-scale analyses, from soma positioning to fine-grained axon tract tracing, and enables cross-comparison across brain structures~\cite{jiang2025multiscale}.

The MICrONS program has pushed these boundaries even further by reconstructing a cubic millimeter of mouse visual cortex at nanometer resolution. Using serial TEM at $\approx$4 nm × 4 nm × 30 nm voxels, MICrONS captured over 120,000 neurons and roughly 0.5 billion synapses and co-registered this with dense two-photon calcium imaging of $\approx$75,000 neurons active during visual stimuli. These co-registered functional and structural volumes provide an unprecedented multimodal connectome, enabling detailed circuit analyses and linking morphology to activity patterns \cite{baeFunctionalConnectomicsSpanning2025a}.

Moreover, integrating radiomic features from fMRI-based Functional Connectivity maps (FC maps) with graph metrics has revealed multi-scale biomarkers that boost diagnostic accuracy. Pirozzi et al. reviewed eleven such studies and found that models combining first-order and texture features from DC, VMHC, ALFF, and ReHo with measures like Regional Radiomic Similarity Networks and rich-club coefficients achieved AUCs > 0.80 in schizophrenia, ADHD, and Parkinson’s—over 0.15 higher than connectomics alone \cite{Pirozzi2025}. This evidence highlights how node-level heterogeneity complements edge-level connectivity and suggests brain-inspired AI and neuromorphic designs should incorporate locally diverse processing units.

The fly datasets released in 2024 show complete, cell-typed wiring diagrams for an adult \textit{Drosophila} brain, allowing direct comparisons of morphology, connectivity, and neurotransmitter profile across three hemispheres \cite{schlegelWholebrainAnnotationMulticonnectome2024}. Follow-up computation work shows that stimulus-specific activation patterns can be reproduced with very simple neuron models once the authentic graph is preserved, and that even a 1\% random rewiring is enough to erase these patterns \cite{zhangNetworkStructureGoverns2025}. These studies provide evidence that network topology, rather than sheer neuron count or biophysics, governs emergent function. 

Together, these resources advance our ability to decode the structural logic of brain function and offer a rich substrate for AGI systems to learn from biologically grounded connectivity patterns. As digital neuroanatomy scales, it sets the stage for integrative models that combine molecular, anatomical, and dynamical information, providing the architecture upon which brain-inspired computation and neuromorphic design can be based.

\subsection{Large-Scale Multilevel Brain Simulations}
Large-scale multilevel brain simulations seek to bridge the gap between detailed cellular models and whole-brain dynamics by integrating neuronal biophysics, microcircuit reconstruction, and macroscopic network simulations into a unified framework. Early proof-of-concept work by Izhikevich and Edelman demonstrated that a one-million-neuron thalamocortical network model could reproduce emergent oscillations and synchrony patterns observed in vivo, providing a first glimpse of how large populations give rise to system-level rhythms \cite{Izhikevich2008}. Building on such foundations, Markram et al. reconstructed and simulated a neocortical microcircuit of the rat somatosensory cortex—31, 000 neurons interconnected via 37 million synapses—revealing how the precise connectivity motifs shape network responsiveness and adaptation as part of the Blue Brain Project \cite{Markram2015}. 
At an even higher level of abstraction, The Virtual Brain platform uses individual MRI-derived connectomes coupled to neural mass models to simulate whole-brain functional dynamics, capturing both healthy resting-state networks and clinically relevant alterations in disorders such as epilepsy and Alzheimer’s disease \cite{SanzLeon2015}. Proponents argue that multiscale simulations like these are indispensable for testing mechanistic hypotheses: they allow systematic, multimodal validation against electrophysiological recordings, neuroimaging data, and behavioral outputs to probe how molecular or synaptic perturbations percolate up to affect cognition and behavior \cite{Einevoll2019}.
Taken together, these efforts illustrate a multilevel roadmap—from ion-channel kinetics and synaptic conductances, through reconstructed microcircuits, to whole-brain network dynamics—that promises not only to deepen our scientific understanding of brain function, but also to guide clinical interventions (e.g., personalized neuromodulation) and inspire next-generation neuromorphic AI architectures.
Looking ahead towards the future direction, continued advances in multimodal brain mapping, machine learning-driven parameter tuning, and standardized model sharing are expected to further refine the predictive power and accessibility of large-scale brain simulations. When these tools mature, they will hold the potential not only for academic discovery but also for translational applications, such as patient-specific brain modeling for personalized treatment strategies in neuropsychiatric conditions.

\subsection{Brain Mechanical Simulations}
Mechanical simulations have emerged as a vital approach for elucidating the complex processes underpinning brain development and injury, offering a reproducible, cost-effective complement to experimental and imaging studies \cite{tallinen2016growth}. During brain development, particularly between 16 and 40 post-menstrual weeks, the cortex undergoes rapid volumetric expansion and intricate folding, processes influenced by mechanical interactions within the brain tissue and between the brain and its surrounding structures \cite{zhang2020cortical}. Computational models, such as finite element methods (FEM), have been widely used to simulate the initiation and evolution of cortical folding based on mechanisms including axonal tension, differential growth, and mechanical constraint from the skull and cerebrospinal fluid \cite{razavi2021mechanism, chavoshnejad2023integrated}. Many early simulations assumed homogeneous growth patterns, such as isotropic or tangential expansion, to model cortical development \cite{zhang2017mechanisms,zhang2016mechanism}. More recent studies have introduced region-specific, heterogeneous growth profiles, motivated by experimental observations of spatiotemporal variability in brain expansion \cite{hou2025role,hou2024automated,hou2025exploring,filla2024accuracy}. Additionally, to better capture the anisotropic mechanical behavior of brain tissue, fiber-reinforced models incorporating axonal orientation have been developed, allowing simulations to reflect the influence of white matter tracts on cortical morphology \cite{chavoshnejad2021role}. Parallel to continuum-scale approaches, emerging cell-based models have also been proposed, which represent brain tissue as a dynamic collection of interacting cells \cite{zarzor2021two}. These studies aim to capture the microscale biological processes, such as differential proliferation and local mechanical interactions, that contribute to macroscopic cortical folding \cite{wang2022orchestrated}. Together, these advancements provide a more comprehensive and biologically grounded framework for understanding the mechanical regulation of brain development. Similarly, in the context of traumatic brain injury (TBI), mechanical simulations play a pivotal role in advancing our understanding of injury mechanisms, progression, and potential therapeutic strategies.  

The brain’s heterogeneous and viscoelastic nature, coupled with its confined anatomical environment, leads to complex stress and strain distributions during impact events \cite{hou2025mechanical}. FEM incorporating high-fidelity anatomical structures and material properties allow researchers to simulate different injury scenarios, predict regions of high mechanical vulnerability, and assess the biomechanical correlates of tissue damage \cite{wu2022interdisciplinary}. Through the integration of imaging-derived data and biologically informed material models, mechanical simulations not only deepen our mechanistic understanding of brain development and injury but also offer a promising pathway toward designing targeted interventions and personalized medicine approaches.

Neurodegenerative diseases, such as Alzheimer's disease (AD), involve complex interactions between biochemical processes and mechanical alterations within the brain. Unlike the relatively rapid events of brain development or acute traumatic injuries, the progression of AD unfolds over extended periods, characterized by the accumulation and spread of misfolded proteins like amyloid-beta (A$\boldsymbol{\beta}$) and tau \cite{Braak1991-hv}. These proteins propagate through the brain in a prion-like manner \cite{Jucker2018-oq}, leading to neuronal dysfunction and death, and ultimately resulting in brain atrophy \cite{Chetelat2010-ae,Scheckel2018-es}. To model these intricate processes, researchers have employed multiphysics approaches that integrate biochemical kinetics with mechanical deformation \cite{Schafer2019-tf,Pederzoli2024-bt}. For instance, the Fisher-Kolmogorov (FK) equation has been utilized to simulate the spatiotemporal spread of pathological proteins, capturing the nonlinear dynamics of their propagation \cite{Weickenmeier2018-ym}. By incorporating diffusion tensor imaging (DTI) data, these models can account for the anisotropic diffusion along neural fiber tracts, reflecting the directional preference of protein spread in white matter \cite{Weickenmeier2019-fy}. Furthermore, coupling these biochemical models with mechanical simulations allows for the assessment of tissue deformation resulting from protein accumulation and neuronal loss. Such integrated models have successfully reproduced characteristic patterns of brain atrophy observed in AD \cite{Schafer2019-tf, Pederzoli2024-bt}, providing insights into the disease's progression and potential therapeutic targets. Despite these advancements, challenges remain in accurately capturing the heterogeneity of brain tissue properties and the multifaceted nature of disease progression. Future research should focus on enhancing model fidelity by incorporating region-specific mechanical properties, nonlinear tissue behavior, and patient-specific data. Additionally, exploring the role of mechanical stress in inducing cellular apoptosis could offer new perspectives on the mechanisms underlying neuro--degeneration \cite{Quan2014-er,Valon2019-ag}. 

\subsection{Brain-Inspired Computing Concepts}
Artificial intelligence is inherently inspired by biological intelligence. Naturally, the development of machine intelligence would rely on the human brain as both a source of inspiration and a benchmark to reach and eventually surpass, since it is the only example of higher intelligence in the natural world. There are many examples of brain-inspired computing concepts, with some even dating back to the origins of AI research. Neural networks, the architectures driving the recent successes behind deep learning approaches, are clear examples of the potential benefit of utilizing concepts from neuroscience in AI research \cite{mcculloch1943}. 
As neural networks were explored further in the decades following their introduction, many methods employed related concepts from neuroscience to improve their performance. For example, dropout, which is a technique that prevents overfitting by only activating a random set of parameters selected with a given probability, was directly inspired by the stochastic nature of neuron group firing rates \cite{hinton2012}. Similarly, pruning, which optimizes a network’s size by removing its least important parameters, is said to be inspired by the brain’s process of removing weak synaptic connections to further strengthen frequently used pathways during childhood and adolescence \cite{bingham2025, wang2020}. Even beyond neural networks explicitly, neuroscience's influence on AI is undeniable. Reinforcement learning is directly inspired by research in animal learning and conditioning \cite{hassabis2017}, and one of its biggest breakthroughs was its successful integration with deep learning \cite{silver2016}. This shows how the effects of two brain-inspired approaches can compound together to outperform even the best humans in a classical board game like Go. 
In current research, transformer architectures have received increasing recognition due to their ability to account for long-range contextual information in tasks like natural language processing and computer vision \cite{vaswani2017, dosovitskiy2021}. A defining characteristic of the transformer architecture is the attention mechanism, which was inspired by attentional mechanisms in the human visual system \cite{mnih2014}. Despite originating in the mid-twentieth century, brain-inspired concepts continue to offer potential for advancement in the present day. As we look to the future, there are still many features of the brain that, when simulated appropriately, may further push the capabilities of current AI, including efficient learning, transfer learning, and an intuitive understanding of the physical world \cite{hassabis2017}.

\subsection{Emerging Directions in Brain Function and Modeling}

\subsubsection{Single-cell and spatial multi-omics} 
Rapid advances in single-cell sequencing and spatial multi-omics are unravelling the brain’s cellular diversity and molecular architecture at higher resolutions \cite{zhangMolecularlyDefinedSpatially2023,jungSpatialTranscriptomicsNeuroscience2023}. High-throughput single-cell transcriptomics, epigenomics, and spatial omics methods now map the expression profiles and spatial positions of thousands of cell types across whole brains. Notably, the first complete molecular cell atlas of a mammalian brain has been achieved, combining single-cell profiles with in situ spatial localization of those cells’ gene expression patterns \cite{zhangMolecularlyDefinedSpatially2023}. These multimodal datasets reveal how neuronal and glial subtypes are arranged and interact in specific circuits, linking gene expression to connectivity and function. Such insights inform neuromorphic modeling by providing biologically grounded “parts lists” and blueprints for circuit architectures, enabling artificial neural networks that more faithfully emulate the rich cell-type heterogeneity and organization of biological brains \cite{jungSpatialTranscriptomicsNeuroscience2023}. Integrating multi-omics data into brain-inspired AI may also guide the design of neuronal units or subcircuits with specialized biochemical and functional properties, bridging molecular neuroscience and neuromorphic engineering.

\subsubsection{Whole-Brain Functional Imaging} 
Emerging imaging techniques now capture brain activity at the whole-organ scale with fine detail, illuminating how distributed neural dynamics give rise to cognition. Next-generation microscopy and recording methods, such as fast volumetric two-photon imaging or functional ultrasound and ultra-high-field fMRI, can monitor neural activity across large brain volumes in real time \cite{hoffmannBlazedObliquePlane2023,vertTranscranialBrainwideFunctional2025}. For example, innovative light-sheet and oblique plane microscopy approaches have achieved cellular-resolution recording across an entire adult vertebrate brain, linking fine-scale neuronal activity to global network states \cite{hoffmannBlazedObliquePlane2023}. In parallel, noninvasive modalities like transcranial functional ultrasound imaging now map hemodynamic responses throughout the rodent brain with spatiotemporal resolution bridging the gap between electrophysiology and fMRI \cite{vertTranscranialBrainwideFunctional2025}. These whole-brain datasets reveal emergent phenomena (e.g., wave propagation, global oscillations, network-level phase coordination) that inform theories of brain-wide integration. 

\subsubsection{Neuromodulation Models} 
Neuroscience is increasingly illuminating how neuromodulators (such as dopamine, serotonin, acetylcholine, and norepinephrine) globally reconfigure brain circuits, leading to new computational models of neuromodulation in both biological and artificial systems. Recent experiments have enabled simultaneous recording and manipulation of multiple neuromodulatory systems, revealing how their interactions shape learning and behavior \cite{cardozopintoOpponentControlReinforcement2025}. For instance, dopamine and serotonin signals in the striatum have been shown to act in opponent but complementary ways to drive reinforcement learning, a finding now incorporated into refined models of reward processing \cite{cardozopintoOpponentControlReinforcement2025}. The broader lesson is that neuromodulators dynamically tune network excitability, plasticity, and effective connectivity, allowing one hard-wired circuit to perform multiple functional roles depending on context \cite{marderNeuromodulationNeuronalCircuits2012}. Computationally, this has spurred interest in neuromodulatory deep learning algorithms that adjust learning rates, gating, or network parameters in response to global feedback signals, mirroring the brain’s use of chemicals to switch between motivational or attentional states. 

\subsubsection{Cognitive Map Theory} 
The concept of the “cognitive map” — an internal spatial representation of relationships in an environment — has evolved from a theory of hippocampal function in navigation to a broader principle of how brains organize knowledge. Mounting evidence indicates that neural circuits (especially in the hippocampus, entorhinal cortex, and connected networks) encode not only physical space but also abstract variables and task structures in a map-like format \cite{behrensWhatCognitiveMap2018}. These brain regions appear to construct geometries or manifolds where conceptually related experiences are positioned nearer to each other, enabling generalization and flexible inference. Recent human fMRI studies, for example, demonstrate that the hippocampus and orbitofrontal cortex collaboratively map out high-dimensional abstract task spaces during learning \cite{qiuFormingCognitiveMaps2024}, much as place cells and grid cells map environments. This expanded cognitive map theory posits that the brain stores a universal chart of relationships (spatial, temporal, social, etc.) to support planning and prediction beyond literal navigation. Inspired by these findings, AI researchers are incorporating cognitive-map-like representations into artificial agents, allowing knowledge to be encoded in relational graphs or vector spaces that facilitate transfer learning and imagination-based planning.

\subsubsection{Energy Constraints in Neural Coding} 
A crucial but often underappreciated driver of neural computation is the severe energy budget under which biological brains operate. The human brain consumes on the order of only 20 watts of power, which imposes strict constraints on neural coding strategies and information processing \cite{attwellEnergyBudgetSignaling2001}. To maximize efficiency, brains have evolved sparse and predictive coding schemes that minimize energetically costly action potentials and synaptic activity while still transmitting critical information. Theoretical and computational studies now underscore energy optimization as a unifying principle: networks trained to minimize their energy usage naturally develop predictive coding architectures, automatically reducing redundancy and focusing on surprising inputs \cite{aliPredictiveCodingConsequence2022}. This connection between metabolic cost and coding strategy suggests that many features of neural design (such as low firing rates, high synaptic efficacy, and event-driven signaling) arise from pressure to conserve energy. 

\subsubsection{Embodied Sensorimotor Integration} 
In contrast to viewing cognition as disembodied symbol processing, contemporary neuroscience highlights that intelligence is fundamentally embodied, arising from continuous sensorimotor interaction with the environment. Neural circuits seamlessly integrate sensory inputs with motor outputs, even at the earliest stages of processing, indicating that perception and action are deeply interwoven \cite{stringerSpontaneousBehaviorsDrive2019}. For example, large-scale recordings in mice show that “visual” cortex neurons encode not only visual stimuli but also information about the animal’s running, whisking, and other behaviors, suggesting that the brain encodes a holistic sensorimotor state rather than separate modality-specific snapshots \cite{stringerSpontaneousBehaviorsDrive2019}. This perspective aligns with theories of embodied cognition, proposing that higher cognitive processes are grounded in the body’s interactions and the internal simulation of action. Recent work has extended this idea to AI, positing that physical embodiment (or at least virtual embodied experience) is vital for developing general intelligence \cite{barrettMindsMovementEmbodied2024}. Consistent with that view, neuromorphic systems are increasingly integrating perception with motor control loops using event-driven sensors and adaptive control policies, mirroring biological reflexes and feedback cycles. Acknowledging the embodied nature of brain function is steering both neuroscience and AI toward models in which cognition emerges from the closed-loop interplay of brain, body, and environment.
\section{Recent Advances in Large Language Models and Developing AGI}\label{Section 4}

\subsection{Scaling Laws \& Emergence}
Recent years have witnessed remarkable progress in the development of AGI through the scaling of LLMs. The key driver behind these advancements is the empirical discovery of scaling laws, which demonstrate predictable improvements in model performance as a function of increased parameters, dataset size, and computation \cite{kaplan2020scaling}. These scaling trends suggest that model capabilities grow smoothly with scale, provided that training follows optimal compute-efficient pathways.

However, scaling does not entirely enhance performance quantitatively; it also unlocks new qualitative behaviors. As models grow larger, they exhibit emergent capabilities that were absent in smaller counterparts. One of the most striking examples is few-shot learning, where a model, without fine-tuning, can adapt to new tasks simply by observing a few examples embedded in the prompt, without requiring additional training \cite{brown2020language, wang2020generalizing}. The phenomenon of in-context learning indicates that models perform a form of meta-learning internally, solely through inference. A key observation is that these emergent behaviors in large language models often arise abruptly rather than through gradual improvement, typically once the model surpasses certain scale thresholds \cite{wei2022emergent}.
Moreover, understanding the relationship between scaling and emergence remains an active area of research, carrying profound implications for AGI development. If the emergence of general-purpose reasoning continues with scale, then strategically shaping model growth, through architectural choices, training objectives, and data quality, may offer a compelling direction for developing increasingly general and capable models. These insights into scaling laws and emergent abilities have significantly influenced the direction of AGI research.

\subsection{Architecture \& Algorithmic Innovations}
\subsubsection{Reinforcement Learning with Human Feedback}
Reinforcement Learning with Human Feedback (RLHF) is a method that aligns machine learning models, especially LLMs, with human intent by integrating human feedback directly into the training loop. Wu et al. highlighted that predefined training data often struggle with data scarcity, outdated datasets, and limited interpretability \cite{WU2022364}. Rather than relying on predefined data and insufficient reward functions, RLHF allows models to learn from human judgments about which outputs are preferred. Ouyang et al. applied RLHF to fine-tune GPT-3 models into a more instruction-following variant called InstructGPT and showed that even a 1.3B parameter InstructGPT model outperforms the much larger 175B GPT-3 model in terms of helpfulness, truthfulness, and reduced toxicity, all while using far fewer parameters \cite{ouyang2022traininglanguagemodelsfollow}. Bai et al. included harmlessness training and explored the tension between helpfulness and harmlessness when aligning LLMs with human values using RLHF \cite{bai2022traininghelpfulharmlessassistant}. Their findings showed that over-optimizing for harmlessness can lead to overly cautious models, while a balanced training approach improves both safety and utility. Moving forward, interactive and discursive feedback mechanisms—designed to surface nuanced disagreements—along with advanced aggregation methods such as multi-objective optimization and weighted feedback aggregation, can be integrated into current RLHF frameworks \cite{GonzlezBarman2025}. This integration may lead to more robust and context-sensitive outcomes while ensuring that feedback is not disproportionately skewed toward any single perspective.
Additionally, RLHF can significantly enhance reward shaping in neuromorphic agents, such as SNN-based robots or embedded systems, especially in sparse or noisy environments. By incorporating guided human feedback, these systems can learn desirable behaviors more efficiently, with fewer interactions, while also avoiding risky or undesirable states. This is particularly valuable for real-world neuromorphic applications where traditional reinforcement signals are difficult to define or unreliable.

\subsubsection{RLAIF and d-RLAIF}
Though RLHF proves effective at aligning model output to human preferences, collecting high-quality human data is incredibly time consuming and costly, bottlenecking scalability. Ultimately, RLHF performance becomes more stale post-training, and alignment performance decreases over time. LLMs show high alignment with human preferences as a baseline, making it possible to collect preference data from other LLMs for alignment. This is known as Reinforcement Learning from AI Feedback (RLAIF), a scalable and cost-effective alternative to RLHF. 
RLAIF can achieve performance comparable to, and in the case of harmless dialogue generation, even superior to RLHF across various tasks including summarization and helpful dialogue generation \cite{pulari2024enhancing}. A key contribution to alignment research is the finding that RLAIF facilitates a form of “self-improvement,” where an AI model providing feedback can be the same size as, or even the exact same checkpoint as, the policy model being trained, yet still yield substantial improvements over supervised fine-tuned baselines. Furthermore, the introduction of direct-RLAIF (d-RLAIF), which obtains reward signals directly from an off-the-shelf LLM during the reinforcement learning process without explicit reward model training, offers an even more streamlined approach that matched or outperformed canonical RLAIF. These results are pivotal for the field of model alignment, as they highlight a viable path towards reducing reliance on expensive and time-consuming human preference labeling, thereby making the process of aligning large language models with human values and intentions more scalable and accessible.

Several optimization methods have also been explored in order to make RLHF more efficient and effective. This includes Direct Preference Optimization (DPO), Contrastive Preference Optimization (CPO), Simple Preference Optimization (SimPO), and their variants—that refine or streamline the traditional preference-based training pipeline \cite{rafailov_direct_2024, xu_contrastive_2024, meng_simpo_2024, xu_fe1ixxucpo_simpo_2025, xiao_cal-dpo_2024, liu_dora_2024}. 
While a detailed review is beyond the scope of this work, these techniques are noteworthy for their potential to improve the scalability, efficiency, and alignment quality of large language models. In particular, methods that reduce reliance on explicit reward models or simplify optimization objectives may be especially relevant for neuromorphic or hardware-constrained systems, where computational and memory resources are at a premium. As such, ongoing progress in preference optimization not only advances alignment, but also supports the broader integration of algorithmic and architectural innovations across both conventional and brain-inspired AI.

\subsubsection{Further innovations in model alignment}
While reference‑free alignment methods and parameter‑efficient adapters such as DoRA have demonstrated notable performance gains over the traditional RLHF pipeline and full‑parameter fine‑tuning, their cost‑effectiveness on smaller language models remains under‑explored, particularly when multiple strategies are combined. To clarify this open question, we survey in Section 4.3 the recent preference‑optimization algorithms: Direct Preference Optimization (DPO), Contrastive Preference Optimization (CPO), Simple Preference Optimization (SimPO), Calibrated DPO (Cal‑DPO), and the hybrid CPO‑SimPO. We discuss how each modifies the original RLHF objective (e.g., eliminating a reward model in CPO/SimPO or rescaling reward estimates in Cal‑DPO). In parallel, we outline how Weight‑Decomposed Low‑Rank Adaptation (DoRA) can cut memory and compute overhead by restricting updates to compact weight decompositions, making it an attractive companion to the above alignment techniques for resource‑constrained deployments.

\subsubsection{System-level Co-design}
Zhao et al. \cite{zhao2025deepseek} show through DeepSeek-V3 that aligning very large MoE models is practical only when the model path and cluster path are tuned together. FP8 kernels plus Multi-head Latent Attention halve activation-memory traffic; an eight-plane two-layer fat-tree, paired with "node-limited" expert routing, keeps all-to-all bandwidth high even on 400 Gb NICs; and dual micro-batch pipelining hides those communications behind attention/MoE compute via GPUDirect Async. On an even larger dense model, Jiang et al.'s MegaScale infrastructure \cite{jiangMegaScaleScalingLarge2024} reaches 55\% MFU on 12,288 GPUs by hierarchically composing pipeline, data, and tensor parallelism, then driving per-step congestion telemetry into a feedback controller that retunes NCCL ring layouts and all-reduce chunking on the fly. Alibaba HPN \cite{qianAlibabaHPNData2024} addresses the bandwidth bottleneck by reshaping the data‑center network itself: a lean two‑tier design with twin top‑of‑rack switches, rail‑based 1,024‑GPU “segments,” and carefully separated routing paths prevents load‑balancing clashes and rack‑level failures; the resulting 15,000‑GPU pod delivers a 14.9\% speed‑up on Llama‑style training while letting each server saturate its 3.2 Tb/s of network bandwidth.Finally, Fire‑Flyer AI‑HPC \cite{anFireFlyerAIHPCCostEffective2024} shows that a 10,000‑GPU PCIe cluster can achieve about 80\% of DGX‑A100 per‑GPU throughput while delivering equivalent overall work at roughly half the hardware cost and 40\% lower energy, by off‑loading all‑reduce to CPUs (HFReduce), overlapping compute/communication (HaiScale), and running compute and storage over a shared two‑tier InfiniBand fabric.

Together, these studies emphasize that future gains in alignment and scaling will come as much from precision formats, topology choices, and traffic‑shaping firmware as from innovations in loss functions or preference data.

\subsubsection{Retrieval-Augmented Generation}
Large language models have recently been enhanced by retrieving and integrating external knowledge at generation time, a paradigm known as retrieval-augmented generation (RAG). A seminal architecture by Lewis et al.~\cite{lewis2020retrieval} introduced RAG for knowledge-intensive tasks, combining a neural retriever with a sequence-to-sequence generator trained end-to-end. This approach enabled models to draw facts from Wikipedia in producing answers, achieving state-of-the-art results on open-domain question answering and demonstrating that even relatively compact generators can outperform much larger parametric models when augmented with relevant documents. Subsequent research improved the retriever-generator synergy: Izacard and Grave~\cite{izacard2020leveraging} proposed a Fusion-in-Decoder model that feeds multiple retrieved passages into the encoder-decoder, allowing the generator to aggregate evidence from many sources. This yielded substantial gains on benchmarks like Natural Questions, highlighting the benefit of richer retrieval inputs. Borgeaud et al.~\cite{borgeaud2022improving} further scaled RAG in the RETRO system, which retrieves from a colossal corpus (trillions of tokens) during generation. Notably, RETRO’s 7.5B-parameter model was shown to match the performance of language models an order of magnitude larger without retrieval, underscoring how external knowledge access can amplify a model’s effective capacity.
More recent efforts integrate retrievers even more tightly with LLMs and broaden RAG’s applications. Izacard et al.~\cite{izacard2022few} introduced \textit{Atlas}, a retrieval-augmented language model pre-trained and fine-tuned for knowledge-intensive tasks under few-shot settings. Atlas achieved state-of-the-art performance on several knowledge-heavy benchmarks with limited training examples, illustrating efficient knowledge integration and adaptation via RAG. Beyond question answering, RAG has been applied to open-ended generation tasks such as knowledge-grounded dialogue. For instance, Shuster et al.~\cite{shuster2021retrieval} incorporated a real-time document retriever into a conversational agent, which substantially reduced factual hallucinations by grounding responses in retrieved evidence. Their RAG-based dialog system attained leading results on knowledge-intensive conversation benchmarks, confirming the paradigm’s value in enhancing the accuracy and reliability of generated text. These developments have established RAG as a powerful approach for leveraging external knowledge, enabling models to produce more informative and up-to-date content across a range of generation tasks.

\subsection{Benchmarks \& Evaluating Intelligence}
Massive Multitask Language Understanding (MMLU) covers 57 domains, ranging from law and medicine to abstract mathematics and history, challenging models to demonstrate not only factual recall but also the ability to reason across diverse fields of human knowledge. Notably, early large models, such as GPT-3, faced significant difficulty in achieving performance far above chance. However, by 2023, models like GPT-4 exhibited near-expert-level human performance, marking a significant improvement in the models' reasoning and knowledge across multiple areas \cite{hendrycks2020measuring}.

BIG-Bench (Beyond the Imitation Game), emphasizes emergent reasoning and generalization. This benchmark suite, crafted by over 450 contributors, includes more than 200 tasks covering symbolic reasoning, common sense, childhood development puzzles, and programming. Importantly, BIG-Bench revealed a phenomenon now central to AGI discourse: certain capabilities in LLMs emerge only beyond a critical model scale\cite{srivastava2022beyond}. As these tasks became tractable to GPT-4-class systems, researchers introduced BIG-Bench Hard (BBH) and BIG-Bench Extra Hard (BBEH) to challenge these models further \cite{kazemi2025big}.

TruthfulQA tests a model’s resistance to misinformation and cognitive traps. It asks questions designed to elicit common human errors, providing insight into a model’s internal coherence and critical reasoning \cite{lin2021truthfulqa}.
 WinoGrande evaluates nuanced pronoun resolution—a task requiring deep contextual understanding—and HumanEval provides programmatic challenges that test step-by-step logical reasoning. A recurring issue with these benchmarks is saturation. Once a model consistently achieves 90–95\% accuracy, the benchmark loses diagnostic power \cite{sakaguchi2021winogrande}.

An emerging trend is the use of human-standardized tests, like the SAT, GRE, Bar Exam, and medical licensing exams, to evaluate LLMs. These tests, though not originally intended for AI, now serve as informal AGI barometers. GPT-4, for instance, ranks in the top 10\% of test-takers for the Uniform Bar Exam and top percentiles in various AP subjects \cite{martinez2024re}. This repurposing of human exams reflects a shift in how intelligence is assessed: the goal is not merely to solve domain-specific tasks but to demonstrate integrated, generalized reasoning across varied contexts.

Alan Turing’s imitation game asked whether a machine could convincingly simulate human behavior in a conversation. While foundational, the Turing Test has limitations—it favors linguistic mimicry over genuine problem-solving. A more formalized approach is a mathematical definition of intelligence as the ability of an agent to perform well across a weighted distribution of computable environments. Although not directly computable, this theory highlights generality as the hallmark of intelligence and remains a touchstone in AGI literature\cite{legg2007universal}.

François Chollet introduced the concept of “skill-acquisition efficiency” in his landmark paper On the Measure of Intelligence. Chollet distinguishes between mere skill at a task and the capacity to acquire new skills with minimal prior knowledge. To evaluate this, he created the Abstraction and Reasoning Corpus (ARC)—a benchmark comprising visual puzzles that require abstract reasoning and generalization from a handful of examples \cite{website2025ARC-AGI}.

The following table shows the most recent and prominent benchmarks and tools used to evaluate intelligence in AGI and LLMs.

\begin{longtable}{@{}p{3cm} p{1cm} p{6cm} p{4cm}@{}}
\caption{Benchmarks for LLM Evaluation}\label{tab:benchmarks}\\
\toprule
\textbf{Tool / Benchmark} & \textbf{Year} & \textbf{Summary} & \textbf{Category} \\
\midrule
\endfirsthead

\multicolumn{4}{@{}l}{\small\itshape Continued from previous page}\\
\toprule
\textbf{Tool / Benchmark} & \textbf{Year} & \textbf{Summary} & \textbf{Category} \\
\midrule
\endhead

\midrule
\multicolumn{4}{@{}r}{\small\itshape Continued on next page}\\
\endfoot

\bottomrule
\endlastfoot

MMLU & 2021 & Massive Multitask benchmark for factual knowledge and reasoning over 57 subjects. & General Multitask Evaluation \\
BIG-Bench & 2022 & Community-built 204-task benchmark to evaluate emergent capabilities in LLMs. & General Multitask Evaluation \\
BIG-Bench Hard (BBH) & 2022 & Subset of especially challenging BIG-Bench tasks. & General Multitask Evaluation \\
BIG-Bench Extra Hard (BBEH) & 2025 & More difficult variant of BBH, designed to push limits of GPT-4-class models. & General Multitask Evaluation \\
TruthfulQA & 2022 & Evaluates truthfulness and resistance to misinformation in LLMs. & Commonsense Reasoning \& Truthfulness \\
WinoGrande & 2020 & Pronoun resolution task testing commonsense reasoning. & Commonsense Reasoning \\
PIQA & 2020 & Physical interaction and commonsense reasoning tasks. & Commonsense Reasoning \\
HellaSwag & 2019 & Tests contextual understanding and story completion. & Commonsense Reasoning \\
GSM8K & 2021 & Grade-school math word problems for step-by-step reasoning. & Mathematical Reasoning \\
MATH & 2021 & High-school to competition-level math problems. & Mathematical Reasoning \\
HumanEval & 2021 & Evaluates code generation and functional correctness. & Programming Ability \\
ARC (Abstraction and Reasoning Corpus) & 2019 & Visual analogical reasoning tasks designed to evaluate generalization and skill acquisition. & Theoretical Evaluation \\
AgentBench & 2024 & Tests LLM-based agents in interactive environments for planning and adaptation. & Agentic \& Interactive Evaluation \\
MMLU-Pro & 2024 & Perturbed variant of MMLU to assess robustness under rephrasing/noise. & Robustness Evaluation \\
HELM (HolisticEval) & 2022 & Multi-dimensional evaluation including bias, calibration, and fairness. & Robustness \& Ethical Evaluation \\
PsychoBench & 2024 & Psychological profiling of LLMs using clinical personality scales. & Psychological \& Human-likeness Evaluation \\
AGENT & 2021 & Evaluates Theory of Mind and goal reasoning in LLMs. & Cognitive Psychology Inspired \\
Animal-AI Olympics & 2019 & Benchmarks inspired by animal cognition experiments. & Animal Cognition \& Embodied AI \\
PsychLab & 2018 & Suite of psychological experiments adapted for reinforcement learning agents. & Cognitive Task Evaluation \\
Brain-Score & 2020 & Neural similarity benchmark comparing AI vision model activations to brain recordings. & Brain-Likeness Evaluation \\
NeuroBench & 2025 & Neuromorphic benchmark for continual learning, vision, motor decoding, and efficiency. & Neuromorphic Computing Evaluation \\
BetterBench & 2024 & Meta-benchmark analysis reviewing benchmark quality and integrity. & Benchmark Design \& Critique \\
LiveBench & 2024 & Dynamic benchmark with secret rotating test sets to prevent overfitting. & Benchmark Integrity \& Evolution \\
\end{longtable}

Numerous studies have utilized established benchmarks as depicted in Table \ref{tab:benchmarks} to evaluate large language models (LLMs) across diverse fields such as science, mathematics, and logical reasoning. Zhong et al. provided a comprehensive survey of the OpenAI o1-preview model, assessing its reasoning capabilities across various domains including code generation, medical knowledge, radiology, robotics, educational measurements, and art education \cite{zhong2024evaluationopenaio1opportunities}. Their findings indicated that o1-preview exhibited outstanding performance in advanced reasoning, domain-specific knowledge, and scientific and medical reasoning. However, despite its impressive capabilities, o1-preview demonstrated limitations in handling highly abstract logical puzzles and adapting effectively to dynamic real-time situations.

Following the significant success of o1-preview, OpenAI continued developing more sophisticated reasoning models, such as the o3 model. According to Myrzakhan et al., the o3 model employs a reflective approach, pausing and adjusting its reasoning processes before providing a final response. This human-like reasoning enables more nuanced and adaptive problem-solving, particularly in highly complex scenarios \cite{myrzakhan2024openllmleaderboardmultichoiceopenstylequestions}.

In evaluating semantic comprehension and knowledge utilization for inference, the Knowledge-Oriented Language Model Evaluation (KoLA) benchmark emphasizes rigorous assessment of language models' capabilities \cite{yu2024kolacarefullybenchmarkingworld}. Jahin et al. conducted comparative analyses of various LLMs across three mathematics datasets, specifically targeting mathematical reasoning \cite{jahin2025unveilingmathematicalreasoningdeepseek}. Their findings underscored that OpenAI’s o3-mini model achieved exceptional performance in mathematical reasoning tasks.

To address the unique obstacles in healthcare-focused LLM development and deployment, Tam et al. introduced QUEST, a comprehensive and practical evaluation framework \cite{Tam2024}. QUEST incorporates five critical evaluation principles: understanding and reasoning, information quality, trust and confidence, expression style and persona, and safety and harm.

Rather than relying solely on static datasets or human evaluation methods, Li et al. proposed DeepEval, an innovative evaluation method emphasizing deep interactions among multiple LLMs within carefully constructed scenarios that simulate realistic interactive environments\cite{li2023staticdatasetsdeepinteraction}. Their experimental results demonstrated that GPT-4 consistently outperformed peer models, achieving state-of-the-art performance across various interactive evaluation tasks. Chang et al. categorized LLM evaluation methodologies into automatic evaluation and human evaluation, highlighting the absence of conclusive evidence favoring any single evaluation protocol or benchmark universally \cite{changSurveryEval}. Each approach exhibits distinct characteristics and focal points, underscoring the importance of diverse evaluation strategies.

\subsection{Brain-Inspired Autonomous Agent Systems}
Brain-inspired autonomous agent systems integrate insights from cognitive neuroscience to endow artificial agents with modular architectures that parallel human brain functions such as perception, memory, and decision-making \cite{liu2025advances}. These systems typically decompose agent cognition into interacting modules, such as memory, world modeling, reward processing, and emotion-like regulation, that map onto analogous brain regions to improve robustness and adaptability in dynamic environments \cite{taniguchi2021hippocampal}.
Neural circuit policies (NCPs) represent one influential approach, using spiking neuronal modules inspired by central pattern generators to produce interpretable control policies that are both robust and energy-efficient \cite{lechner2020neural}. In parallel, spiking neural networks (SNNs) and neuromorphic computing hardware have been extensively investigated to enable real-time, event-driven processing for perception and control in autonomous robots \cite{davies2018loihi}. For example, neuromorphic perception frameworks leverage asynchronous event cameras and hippocampus-inspired spatial mapping to achieve low-latency navigation in unstructured environments \cite{casanueva2023bioinspired}.
Complementary bio-inspired methods employ Hebbian-like local learning rules and attractor dynamics to support adaptive behaviors and resilience to sensor noise \cite{oja1982simplified}. Applications of these brain-inspired autonomous agent systems span robotic manipulation, mobile navigation, and autonomous driving, demonstrating improved energy efficiency, interpretability, and adaptability compared to conventional AI controllers \cite{zeng2019learning}.
Despite promising results, challenges remain in scaling these architectures to complex tasks, integrating multi-scale brain mechanisms, and ensuring transparent interpretability of learned behaviors \cite{taniguchi2021hippocampal}. Future research directions include unifying cognitive and neural inspirations within hybrid frameworks, leveraging large-scale brain connectivity data to refine agent modules, and co-designing neuromorphic hardware and learning algorithms for on-board autonomy \cite{taniguchi2021hippocampal}.

\subsection{Limitations \& Research Gaps in Contemporary AI}

\subsubsection{Embodiment}
Despite recent advances, LLMs still face important cognitive limitations. A key gap is their lack of embodiment. Human intelligence develops through sensory and physical interaction with the world, but LLMs learn only from text, leaving their understanding detached and abstract \cite{chemero2023llms}. As a result, they cannot perceive or act in the physical world. Even multimodal models, which process images or videos, do not truly engage with their environment, limiting their grasp of meaning \cite{jones2024multimodal,agarwal2025cosmos}. This lack of grounding can lead to errors or hallucinations, outputs sound plausible but are factually incorrect or misleading \cite{xu2024hallucination}.

\subsubsection{Causal Reasoning}

LLMs also struggle with causal reasoning. While they can describe correlations, they lack mechanisms to model cause-and-effect relationships \cite{yu2025causaleval}. Without a grounding in physical experience or the ability to simulate interventions, their reasoning remains surface-level. This limits performance on tasks requiring inference, counterfactual thinking, or planning beyond learned patterns \cite{zhou2024image}.

\subsubsection{Memory}
Memory is another constraint. Standard LLMs use a fixed context window, restricting long-term retention and learning across interactions \cite{liu2024lost}. They do not update internal knowledge during use and forget earlier content beyond the window. The next-token prediction objective further reinforces short-term pattern matching rather than structured reasoning or goal planning \cite{xu2025towards}. Addressing these gaps may require integrating persistent memory, causal models, and embodied learning frameworks to move beyond current limits.

\subsubsection{Energy}
Recent work highlights that developing and training large language models can release hundreds of metric tons of \ce{CO_2} and consume millions of liters of water: Morrison et al. report that creating a family of models (20 M–13 B parameters) emitted 493 metric tons of \ce{CO_2} and consumed 2.769 million L of water \cite{morrisonHolisticallyEvaluatingEnvironmental2024}.
Inference at scale rapidly becomes a dominant environmental cost. Jegham et al. show that o3 and DeepSeek-R1 each consume over 33 Wh per long prompt, more than 70 × the 0.43 Wh used by a GPT-4.1 nano query, and that processing 700 million such queries daily would use as much electricity as 35,000 U.S. homes \cite{jeghamHowHungryAI2025}.
Even with state-of-the-art optimizations, energy demands remain high. Fernandez et al. demonstrate that applying optimizations such as decoding strategies, batch-size choices, and hardware configurations can reduce total inference energy by up to 73\% compared to unoptimized baselines \cite{fernandezEnergyConsiderationsLarge2025}. Despite efficiency gains, the aggregate power draw of LLMs poses a growing threat to climate goals.

\subsubsection{Interpretability}
	Despite their performance, deep neural networks (DNNs) largely remain “black boxes.” Post hoc methods such as gradient-based saliency maps, perturbation analysis (e.g., LIME, SHAP) and example-based explanations seek to explain pretrained models. However, these suffer from low fidelity under distribution shifts and are vulnerable to adversarial perturbations \cite{bilalLLMsExplainableAI2025}. Intrinsic interpretability embeds transparency into the architecture itself through methods such as prototype layers or attention modules, but scalable designs with competitive accuracy are still in their infancy \cite{jiComprehensiveSurveySelfInterpretable2025}. Mechanistic (white-box) dissection techniques can now map individual neurons or micro-circuits to semantic concepts via network dissection, concept-activation vectors and sparse-auto-encoder feature maps, but the field has only recently begun to standardise evaluation. Two recent examples are the MDPI "Pattern Discovery \& Disentanglement" benchmark for clinical NLP \cite{zhouBenchmarkingInterpretabilityHealthcare2025} and the cross-model Mechanistic Interpretability Benchmark (MIB) \cite{muellerMIBMechanisticInterpretability2025}, both of which publish open leaderboards and hold-out test sets. The European GDPR’s “right to explanation” further underscores the urgency of deploying certifiable interpretability solutions for deployed AI systems \cite{kaminskiRightExplanationExplained2019}. 
\subsubsection{Adaptability}
	DNNs achieve impressive accuracy when the deployment distribution matches the training data, but even mild covariate or concept shifts can trigger a loss of plasticity and performance. \cite{dohareLossPlasticityDeep2024}. Unsupervised domain adaptation (UDA) reduces this gap by aligning source and target feature distributions, but standard pipelines still assume full access to the unlabeled target set during adaptation \cite{rostamiUnsupervisedDomainAdaptation2024}. Source-free domain adaptation (SFDA) does away with that assumption, only transferring a frozen source model, yet recent surveys and benchmarks document optimization instability and accuracy drop-offs on harder benchmarks \cite{yuComprehensiveSurveySourcefree2023,yoonEnhancingSourceFreeDomain2024}. Meta-learning methods (e.g., MAML) now frame each source domain as a "task", training episodically to simulate domain shifts and generalize without ever seeing target-domain data, but they still depend on a rich diversity of source tasks and incur hefty bi-level-gradient costs when scaled to high-dimensional, real-time problems \cite{finnModelAgnosticMetaLearningFast2017,khoeeDomainGeneralizationMetalearning2024}. Continual and lifelong learning address ever-shifting data streams with tactics like experience replay, synaptic-level regularization, and adaptive modularity, but the stability-plasticity equilibrium remains unresolved, and architectural factors are now recognized as the major factor \cite{dohareLossPlasticityDeep2024,luRevisitingNeuralNetworks2024}. Taken together, none of these methods fully solves adaptive robustness. The evidence shows that optimization fragility, compute overheads, and the unsolved stability-plasticity dilemma each flare up when deployment realities deviate from lab conditions.
\section{Recent Advances in Neuromorphic Computing and SNNs}\label{Section 5}

\subsection{Next-Generation Neuromorphic Hardware}
Next-generation neuromorphic hardware is intended to overcome the performance and energy constraints of traditional von Neumann systems. These systems combine memory and computation. They are modeled after the structure and function of the human brain, with spiking neurons, analog computation, and event-driven processing. This strategy improves total efficiency. Nonvolatile memory, analog circuits, photonics, memristors, and in-memory computing arrays are some of the fundamental technologies driving this transition. 

Intel’s Loihi 2 is a good example of digital neuromorphic hardware. It has roughly one million spiking neurons and 120 million synapses, designed for event-driven and asynchronous processing \cite{TakingNeuromorphicComputing2021}. Loihi 2 allows for varied programming of synaptic learning principles, such as spike-timing-dependent plasticity (STDP). This makes it ideal for online and edge learning applications. Unlike traditional chips, which rely on synchronized clock signals, Loihi 2 handles spikes only when they occur. This can reduce power usage by up to 100x for some workloads \cite{TakingNeuromorphicComputing2021}. 

IBM's NorthPole combines SRAM-based memory and digital computation in a 2D mesh of compute units. This design provides inference without the need for off-chip memory access \cite{ambrogio2023northpole}. This improves latency and minimizes energy loss during data transfer, which is a significant bottleneck in traditional CPUs and GPUs. According to Ambrogio et al., NorthPole outperformed leading GPUs in energy efficiency for image inference tasks by around 25x. The device has around 250 million transistors, which are structured to imitate the site of computing in the human cortex.

Memristive crossbar arrays utilize a different approach, storing weights in analog resistance memory elements. They do matrix-vector multiplication directly within the array. Memristors are non-volatile and can maintain their state without constant power, making them energy-efficient for long-term storage \cite{ielmini2018memory}. When memristors are organized in crossbar configurations, they can execute multiply-accumulate operations in constant time, O(1). This enables for highly efficient parallel processing that can scale smoothly with the size of spiking or analog neural networks.

Analog neuromorphic chips, such as BrainScaleS and Neurogrid, use neural models to regulate continuous-time electronic circuitry. BrainScaleS, developed at Heidelberg University, is approximately 1,000 times faster than real biological time. This enables high-performance simulation of massive neural networks. Stanford's Neurogrid supports one million neurons and billions of synapses with less than a milliwatt of power \cite{schuman2022survey}. These devices are designed to imitate true brain function, such as leaky integration, spiking, and refractory periods, while using extremely little energy.

Photonic neuromorphic computing uses light instead of electrical signals to communicate and compute. It employs optical components such as microring resonators and Mach-Zehnder interferometers to conduct multiply-accumulate operations at light speed \cite{shastri2021photonics}. These systems provide high bandwidth, low latency, and are resistant to electromagnetic interference. This makes them ideal for processing massive amounts of real-time sensory data. In experiments, photonic devices attained spike rates in the multi-gigahertz region while consuming only femtojoules of energy per operation. This is far more efficient than standard digital hardware.

Each neuromorphic hardware platform targets a different goal. Some aim to cut power use. Others focus on faster processing or real-time learning. Many are designed to scale to large, brain-like networks. The long-term goal is to build machines as efficient and flexible as the human brain. These systems could eventually operate using just a few watts of power, a stark contrast to today’s GPUs and CPUs, which require hundreds of watts to function.

\subsection{Quantum Neuromorphic Platforms}
While conventional neuromorphic computing, inspired by the structure and function of the biological brain, has made significant strides, it faces fundamental hurdles. Scaling these systems to approach the complexity of the human brain, with its approximately $10^{11}$ neurons and $10^{15}$ synapses, presents immense challenges in terms of power consumption, component density, and fabrication complexity \cite{Schuman2017, Davies2018, Markovic2020}. Furthermore, programmability and system integration remain significant obstacles for current classical neuromorphic platforms. These limitations motivate the exploration of radically different computational paradigms.

Quantum neuromorphic computing emerges as a compelling, albeit nascent, research direction aiming to address these challenges by harnessing the principles of quantum mechanics within brain-inspired architectures. The core proposition extends beyond merely accelerating classical neuromorphic models; it seeks to leverage uniquely quantum phenomena such as superposition, entanglement, and quantum tunneling to potentially realize novel computational capabilities and efficiencies \cite{Melkikh2019, Sanz2021}. By encoding neural states in quantum systems (e.g., qubits, quantum harmonic oscillators) and implementing synaptic interactions through controllable quantum processes, this approach could theoretically access exponentially larger state spaces compared to classical bits, potentially enabling more powerful representations or faster processing for certain tasks. Furthermore, quantum effects might offer pathways to ultra-low energy switching mechanisms for neuronal firing and synaptic plasticity, directly tackling the power scaling issue inherent in classical CMOS-based approaches \cite{Yao2020, Grollier2020}. The goal is not simply to scale existing models but to explore whether mapping neural and synaptic dynamics onto controllable quantum systems unlocks fundamentally different and potentially more advantageous computational primitives.

It is important to distinguish quantum neuromorphic computing from the broader field of Quantum Machine Learning (QML). While QML primarily focuses on developing algorithms that leverage quantum mechanics for machine learning tasks, often designed to run on general-purpose quantum computers \cite{Biamonte2017}, quantum neuromorphic computing is fundamentally concerned with the hardware implementation, building physical systems whose quantum dynamics intrinsically mimic neural computation. Naturally, overlaps exist, particularly in the co-design of algorithms and hardware, but the emphasis in quantum neuromorphic computing lies in creating specialized, brain-inspired quantum processing units.

\subsubsection{Physical Platforms for Quantum Neuromorphic Implementation}
The exploration of quantum neuromorphic hardware spans diverse physical systems, each offering a unique set of potential advantages and facing significant technological hurdles. Several leading platforms are currently under investigation:
\begin{itemize}

\item \textbf{Superconducting Circuits:} Leveraging the mature fabrication techniques developed for superconducting quantum computing, this platform utilizes superconducting elements, most notably Josephson junctions (JJs), as the building blocks for artificial neurons and synapses \cite{Shainline2017, Schneider2022, Crotty2010}. JJs, typically formed by sandwiching a thin insulating layer between two superconductors (e.g., Niobium (Nb) or Aluminum (Al)), exhibit rich nonlinear dynamics governed by quantum mechanics. Their phase difference dynamics can be engineered to emulate the spiking behavior of biological neurons, where exceeding a critical current threshold triggers a voltage pulse analogous to an action potential \cite{McDermott2004, Shainline2018}. Synaptic weights and plasticity can potentially be implemented by coupling JJs via superconducting loops or other quantum circuits, allowing for tunable interactions \cite{Shainline2019}.

A defining characteristic of this approach is the requirement for cryogenic operating temperatures, typically in the milli-Kelvin (mK) range, to maintain superconductivity and suppress thermal noise, thereby preserving quantum coherence \cite{Clarke2008}. This necessitates complex and costly infrastructure, including dilution refrigerators and extensive electromagnetic shielding, posing significant challenges for scalability and practical deployment. However, superconducting circuits offer potential advantages in terms of high operational speeds (gigahertz frequencies) and extremely low energy dissipation per switching event due to the absence of electrical resistance \cite{Solgun2015}. Despite progress in demonstrating basic neuronal and synaptic functions, major challenges remain in scaling these systems to large network sizes, managing qubit/neuron connectivity (wiring complexity), mitigating decoherence effects, and developing efficient interfaces to classical room-temperature control electronics \cite{Krantz2019}.

\item \textbf{Photonic Systems:} An alternative strategy employs photons as the carriers of neural information, offering the potential for operation at room temperature and leveraging the high bandwidth and low crosstalk inherent in optical communication \cite{Shastri2021, Feldmann2021}. In photonic neuromorphic schemes, neural spikes are typically represented by brief pulses of light, potentially encoded using various degrees of freedom such as time-bins, wavelength, or polarization. Integrated photonics, using technologies like silicon photonics (SiPh), enables the fabrication of waveguides, modulators, beamsplitters, and detectors on a compact chip, forming the basis for photonic neural networks \cite{Shen2017}.

Implementing synaptic weights often involves components that modulate light intensity or phase, such as Mach-Zehnder interferometers (MZIs) or microring resonators (MRRs) \cite{Tait2017}. Achieving synaptic plasticity (learning) in purely photonic systems remains a challenge, although hybrid approaches integrating phase-change materials (PCMs) or electro-optic materials within photonic circuits are being explored \cite{Feldmann2019}. A fundamental difficulty in photonic neuromorphic computing lies in realizing strong optical nonlinearities efficiently at low power levels, which are essential for implementing neuron activation functions \cite{Miscuglio2020}. While linear operations like matrix multiplication can be performed very efficiently optically, nonlinear activation is crucial for complex computations. Other challenges include losses in photonic circuits, the efficiency and noise characteristics of single-photon sources and detectors (if operating in the quantum regime), and the difficulty in creating compact, persistent memory elements comparable to electronic synapses \cite{Shastri2021}. Nonetheless, the promise of room-temperature operation and speed-of-light processing continues to drive research, including efforts to harness quantum optical effects like squeezing or entanglement for enhanced computation \cite{Miscuglio2022}.

\item \textbf{Quantum Memristive Systems:} The concept of the memristor, a two-terminal device whose resistance depends on the history of charge flow through it, provides a compelling analogue for synaptic plasticity. Quantum memristive systems aim to push this concept into the quantum realm, utilizing quantum mechanical transport phenomena to achieve memristive behavior at the nanoscale \cite{Pershin2011, Wang20202}. Rather than relying on classical ion migration or bulk phase changes, these devices seek to exploit effects like resonant tunneling through quantum dots, Coulomb blockade, spin-dependent transport in magnetic tunnel junctions, or quantum phase transitions in strongly correlated materials \cite{Sanz2021}. The "memory" in such systems could arise from the configuration of discrete quantum states, charge trapping influenced by quantum tunneling, or structural modifications governed by quantum mechanics \cite{Stiegler2011}.

Materials exploration is central to this research direction. For instance, systems based on superconducting junctions incorporating materials like Hafnium Oxide (\ce{HfO2}) might exploit proximity effects or controlled tunneling barriers to achieve memristive switching \cite{Alagoz2015, Zhao2021}. Two-dimensional (2D) materials like Molybdenum Disulfide (\ce{MoS2}) offer another avenue, potentially leveraging electrically controllable defects, layer interactions, or valleytronic properties to create history-dependent conductance modulated by quantum effects \cite{Xiang2020, Li2017, Sangwan2018}. The allure of quantum memristors lies in their potential for extreme miniaturization, potentially leading to ultra-dense neuromorphic circuits, and the possibility of harnessing quantum switching mechanisms for superior energy efficiency compared to classical memristors \cite{Yao2020}. However, this area faces substantial challenges. Fabricating nanoscale devices with reproducible quantum characteristics is difficult, and precisely understanding and controlling the underlying quantum transport phenomena remains an active area of fundamental research. Demonstrating robust, controllable synaptic behavior and integrating these devices into functional neuromorphic circuits with effective learning rules are key future milestones \cite{Grollier2020}.
\end{itemize}
\vspace{-0.5cm}
\subsubsection{Hybrid Quantum-Classical Neuromorphic Architectures}
Recognizing the formidable challenges in building large-scale, fault-tolerant quantum computers and the concurrent limitations of purely classical neuromorphic systems for certain tasks, hybrid quantum-classical architectures have emerged as a pragmatic near-to-mid-term strategy \cite{Preskill2018}. This approach seeks to leverage the respective strengths of both technologies: the scalability and sophisticated development of classical neuromorphic processors (like Intel's Loihi 2 \cite{IntelHalaPoint}) for handling large network structures and efficient processing of sensory data or routine inference, combined with the potential power of smaller, specialized quantum processors acting as co-processors or accelerators for computationally intractable sub-problems \cite{Neukart2017}.

The motivation stems from the understanding that current and near-term quantum devices operate in the Noisy Intermediate-Scale Quantum (NISQ) era, characterized by limited qubit counts, connectivity, and coherence times \cite{Preskill2018}. While potentially powerful, these devices are unlikely to autonomously execute complex, large-scale neuromorphic algorithms soon. Hybrid architectures propose a division of labor: the classical neuromorphic system manages the overall workflow, data flow, and tasks amenable to classical parallelism, while offloading specific, computationally demanding kernels to the quantum co-processor. Such kernels might include solving hard optimization problems arising during learning \cite{Benedetti2019}, performing probabilistic sampling from complex distributions relevant to Boltzmann machines or generative models \cite{Amin2018}, or potentially accelerating tasks involving complex correlations or combinatorial complexity that challenge classical algorithms \cite{Saggio2021}.

A prominent example involves coupling classical systems with quantum annealers \cite{Johnson2011, Adachi2015}. Quantum annealers are specialized quantum devices designed to find the low-energy ground states of specific optimization problems, typically formulated as Ising models or Quadratic Unconstrained Binary Optimization (QUBO) problems \cite{Hauke2020}. Certain machine learning tasks, particularly in unsupervised learning or reinforcement learning, can be naturally mapped onto such optimization problems \cite{Crawford2016, Neukart2017}. For instance, training Restricted Boltzmann Machines (RBMs) or finding optimal policies in certain reinforcement learning scenarios can involve sampling or optimization steps well-suited to quantum annealing hardware \cite{Amin2018, Dunjko2018}. In a hybrid neuromorphic context, the classical chip could structure the learning problem, passing the core optimization step to the annealer, and integrating the result back into the larger network dynamics.

Despite the conceptual appeal, realizing effective hybrid systems entails significant technical hurdles. A major challenge lies in the physical interface between the quantum and classical components, especially if the quantum processor requires cryogenic temperatures (like superconducting circuits or some annealers). Efficiently transferring information across this temperature gradient while minimizing latency and noise is non-trivial \cite{Das2021}. Bandwidth limitations between the classical host and the quantum co-processor can also create bottlenecks, potentially negating any quantum speedup if communication overhead dominates. Furthermore, developing software frameworks and algorithms capable of intelligently partitioning tasks and managing the complex interplay between the classical and quantum parts is crucial for harnessing the potential of hybrid architectures \cite{Kyriienko2021}.
\subsubsection{System Integration Efforts and Future Prospects}
Moving beyond individual device demonstrations, significant research efforts are now focused on integrating quantum components into functional neuromorphic systems. A notable example is the European Quromorphic project, which explicitly aims to build demonstrators of quantum neuromorphic hardware \cite{QuromorphicProject}. This initiative focuses on developing circuits capable of generating quantum-based spikes and constructing small-scale networks using photonic qubits, directly tackling the system-level integration challenges for both superconducting and photonic platforms. Such coordinated projects signal a shift towards engineering viable quantum neuromorphic processors, providing platforms for exploring algorithms and architectures.

The development of quantum neuromorphic hardware does not occur in isolation. It must be viewed alongside the rapid advancements in classical neuromorphic engineering. Systems like Intel's Hala Point, integrating over 1,100 Loihi 2 chips to simulate networks exceeding one billion neurons, demonstrate the remarkable progress in scaling classical approaches \cite{IntelHalaPoint}. This progress is fueled by innovations in semiconductor technology, including 3D chip stacking for increased density, sophisticated on-chip communication networks, and the development of novel materials for implementing more efficient and scalable artificial synapses \cite{Burr2017}. Materials such as phase-change materials like \ce{Ge2Sb2Te5}~(GST) \cite{Kuzum2013, Boybat2018} and ferroelectric materials like Hafnium Zirconium Oxide (\ce{HfZrO2}) \cite{Mulaosmanovic2017} are enabling high--density, low--power synaptic arrays in classical systems. These classical advancements set a high bar for quantum approaches to surpass, but they also offer potentially synergistic technologies. For example, advanced packaging techniques like 3D integration, developed for classical chips, might prove crucial for integrating disparate quantum and classical components in future hybrid systems.

The path towards practical quantum neuromorphic computing faces several fundamental challenges specific to the quantum domain. Maintaining quantum coherence – preserving the delicate quantum states against environmental noise – long enough to perform meaningful computation is arguably the primary obstacle across most platforms \cite{Schlosshauer2007}. Developing effective error correction or mitigation strategies suitable for the analog-like nature of neuromorphic computation, which may differ significantly from standard quantum error correction (QEC) designed for gate-based quantum computers, is another critical research area \cite{Cai2021}. Scaling the number of quantum neurons and synapses while maintaining high connectivity and individual control remains a daunting engineering task. Furthermore, the quantum-classical interface presents persistent difficulties in terms of latency, bandwidth, and energy efficiency \cite{Das2021}. Finally, designing learning algorithms and computational paradigms that truly exploit the underlying quantum dynamics, rather than merely mimicking classical SNNs, is essential to unlock any potential quantum advantage \cite{Melkikh2019, Sanz2021}.

Progress in the field will also depend critically on the development of standardized benchmarks and software tools to objectively evaluate and compare the performance of diverse hardware platforms – classical, quantum, and hybrid \cite{Schuman2022}. Initiatives like NeuroBench aim to provide common tasks and metrics for classical neuromorphic systems \cite{NeuroBench}, but extensions or new benchmarks tailored to assess the unique aspects of quantum neuromorphic computation (e.g., energy efficiency per task, speedup for specific problems, quality of probabilistic sampling) will be necessary.

In conclusion, the long-term vision for quantum neuromorphic computing is ambitious: to harness quantum phenomena and novel materials within brain-inspired architectures, potentially leading to transformative advances in artificial intelligence characterized by unprecedented speed and energy efficiency for specific computational tasks \cite{Yao2020, Grollier2020}. Realizing this vision necessitates sustained, interdisciplinary research spanning quantum physics, materials science, computer engineering, and computational neuroscience. While still in its early stages, the exploration of quantum effects within neuromorphic frameworks represents a fascinating frontier, holding the potential, alongside continued classical innovation, to shape the future of intelligent computation \cite{Markovic2020, Christensen2022}.

\subsection{Physical Neurons \& Memristor Roles}
Understanding the four fundamental electrical quantities—voltage, current, magnetic flux, and charge—is essential to comprehending the function of the memristor in contemporary circuits. Inductors pair magnetic flux and current, resistors pair voltage and current, and capacitors pair voltage and charge. These are the conventional pairings for passive components. Memristors complete this set by pairing magnetic flux and charge, which introduces a new component to circuit design \cite{Chua1971memristor}.

Among the most promising applications of memristors is neuromorphic computing \cite{Wan2022}. Memristors possess the most significant features of neurons and biological synapses. For example, memristors possess synaptic plasticity: their resistance can be reduced or enlarged by electrical activity, in accordance with how the brain remembers and learns \cite{MercedGrafals2016}. Their parallel processing capability by crossbar arrays is akin to brain neural networks \cite{Xia2019}, and their energy efficiency makes them an energy-frugal choice over traditional computing \cite{Zidan2018}.

Physically, memristors are usually built in a sandwich structure: two electrodes surrounding a very thin semiconductor or insulator film \cite{Wan2022}. Doping by certain impurities results in the layer forming oxygen vacancy areas. These must regulate resistance \cite{Samanta2023}. The smaller doped zone provides high resistance, and vice versa. The boundary between the oxygen-rich and oxygen-poor areas depends on the direction of the current, allowing memristors to provide memory storage as well as logic. Memristors are usually realized in crossbar arrays (CAs) in an attempt to provide dense, scalable computational and memory devices \cite{Xia2019}. In extracting data from arrays, a voltage is applied to a specific word line (row), and current is sensed against a corresponding bit line (column). But the setup is prone to sneak-path currents—undesired currents that lead to errors and power dissipation. Later developments, like the Parallel Read Approach suggested by Indian Institute of Technology engineers, solve this issue by dividing the array and using threshold detectors to identify individual memristors more reliably \cite{Samanta2023}.

Another property of memristors is competitive and cooperative behavior when they share a common source of ions \cite{Xia2019}. In cooperative environments, the action of one memristor may induce corresponding action in other memristors that are nearby, mimicking the process of synaptic strengthening in the brain. In competitive environments, a single memristor may hog all the available ions, making other memristors ineffective. These phenomena mimic phenomena that take place in neural networks, especially sense processing and pattern recognition \cite{Wan2022}.

Memristors offer a robust alternative to the traditional von Neumann architecture, separating memory and processing units \cite{Zidan2018}. When separation occurs, the CPU spends an extended period idle waiting for processor-to-memory data transfers, causing the notorious von Neumann bottleneck. Furthermore, conventional memory technologies like DRAM are power-intensive and volatile, requiring constant refreshing and leading to forgotten data when turned off. Since Moore's Law is becoming less effective, new computer paradigms are needed to keep up \cite{Xia2019}.

In this context, memristors offer a new solution: in-memory computing, or having data computed directly in memory arrays \cite{Zidan2018}. Analog memristor computing eliminates the overhead of constant data transfers, significantly reducing energy and latency expenses. Memristors are highly beneficial for accelerating AI, edge computing, and real-time signal processing due to these factors. Compared with DRAM, dense power-efficient non-volatile memory that is Resistive RAM (ReRAM) but memristive is an improved option \cite{MercedGrafals2016}.

A Memristor-Based Processing Unit (MBPU) would include:
\begin{enumerate}
\item A large-scale memristor crossbar array for data storage and analog computation \cite{Xia2019},
\item A voltage/current control system to manage writing, reading, and computation \cite{Samanta2023}, and
\item A low-power CMOS (Complementary Metal-Oxide-Semiconductor) controller for instruction sequencing, dividing data, and I/O interfacing \cite{Zidan2018}.
\end{enumerate}
Because processing and memory are colocated, the MBPU eliminates the von Neumann bottleneck \cite{Zidan2018}. It also natively supports neuromorphic and analog workloads, thereby being a prime candidate for AI workloads at the edge \cite{Wan2022}.

The MBPU even possesses the ability to carry out basic CPU operations. Addition and subtraction can be achieved by adding row currents by Kirchhoff's Current Law \cite{Xia2019}. Multiplication is done by assuming weights in the form of resistances in the memristor array. Upon the provision of the input voltages on the word lines, the output current for each bit line is equal to the dot product of the voltage vector and the column conductance matrix \cite{Samanta2023}.

Besides arithmetic, memristors can facilitate control flow and logic computation with finite state machines implemented by crossbar arrays \cite{Zidan2018}. They employ material implication logic (IMPLY logic), wherein the value of one state updates another, like: If A is 1, update B to 1; otherwise, don't update B. The other Boolean gates (AND, OR, NOT, NAND, NOR, XOR) can be done together by memristor arrays based on their ability to be both a memory device and a logic device \cite{Xia2019}. This greatly exceeds the abilities of modern logic devices. 

A particular logic function can be learned into any row or column of the array. Because memristor arrays are reprogrammable dynamically, they can change functions, like an FPGA, and allow for customizable processing \cite{Wan2022}.

In summary, memristors complete the long-standing missing link of the elementary circuit components by combining magnetic flux and charge \cite{Chua1971memristor}, providing an extra dimension to electronic design. Their unique aspect of merging memory and computation enables brain-like learning and adaptive behavior, which makes them especially attractive to neuromorphic and in-memory computing \cite{Zidan2018}. Memristor-based systems colocate processing and memory, abolish the von Neumann bottleneck, deliver unmatched energy efficiency, and support reconfigurable logic and analog computation \cite{Xia2019}. While problems like sneak-path currents remain \cite{Samanta2023}, successive innovations in materials and circuit technology are rapidly evolving memristors into useful technologies for AI, edge computing, and beyond \cite{Wan2022}.

\subsection{Advances in SNN Algorithms}
SNNs have experienced a surge in algorithmic innovations in recent years, moving beyond basic rate-based approximations to now encompass biologically plausible learning rules, transformer-like architectures, and self-supervised paradigms. 

\subsubsection{Direct Training with Surrogate Gradients}
	Traditional SNNs suffered from non-differentiable spike functions, making gradient-based training challenging. Surrogate gradient methods overcome this weakness by replacing the hard threshold with a smooth proxy during backpropagation. Lian et al. introduced a Learnable Surrogate Gradient (LSG) mechanism that adapts the shape of a surrogate function per layer and dataset, achieving substantial performance gains on large-scale vision tasks such as ImageNet \cite{lianLearnableSurrogateGradient2023}. Complementing LSG, Li et al. introduced Masked Surrogate Gradients (MSG), which update only a sparse subset of weights at each steps. On ImageNet-scale benchmarks, MSG raises top-1 accuracy by about two percentage points while keeping gradients sparse and adding almost no extra computation \cite{liDirectlyTrainingTemporal2024}.  

\subsubsection{Bio-Plasticity Inspired Learning Rules}
	Building on the STDP rule, recent work has explored more expressive bio-physically grounded mechanisms. Ororbia et al. introduced Contrastive Signal-Dependent Plasticity (CSDP), a self-supervised method that leverages local constraint objectives to drive synaptic updates, achieving robust feature learning without labels in spiking circuits \cite{ororbiaContrastiveSignalDependent2024}. At the same time, Daddinounou and Vatajelu devised a Bi-Sigmoid STDP rule that is tailored to spintronic synapses. The update function directly reflects Magnetic Tunnel Junction (MTJ) physics and results in over 90\% accuracy on unsupervised image classification with in-memory on-chip learning \cite{daddinounouBisigmoidSpiketimingDependent2024}. These rules allow for more biological plausibility and hardware efficiency.

\subsubsection{Spiking Transformer Architectures}
The past two years have seen an explosion of spike‑driven Vision Transformers. Spikformer introduced a fully spike‑based self‑attention module that replaces soft‑max with sparse masking and addition, matching ResNet‑50 accuracy on ImageNet in only four timesteps \cite{zhouSpikformerWhenSpiking2022}. Spike‑Driven Transformer pushed this idea further by devising Spike-Driven Self-Attention (SDSA): the Q-K dot-product and soft-max are replaced with a spike-level mask plus column-wise additions, and the residuals are rerouted through membrane potentials. Together, this cuts multiply-accumulate energy by $\approx$ 87 × while still reaching 77.1\% top‑1 on ImageNet‑1K \cite{yaoSpikedrivenTransformer2023}. More recent variants such as QKFormer use a linear‑complexity Q–K attention and a hierarchical pyramid to push top‑1 ImageNet accuracy past 85\% while retaining T=4 inference \cite{zhouQKFormerHierarchicalSpiking2024}. Lateral‑inhibition designs (SpiLiFormer) \cite{zhengSpiLiFormerEnhancingSpiking2025}, one‑step inference models (OST) \cite{ijcai2024p348}, and mixture‑of‑experts variants (SEMM) \cite{Zhou2024} further show that transformer‑style global context can live comfortably inside an event‑driven, energy‑frugal substrate.

\subsubsection{Self-Supervised and Unsupervised SNN Learning}
Unlabeled data are now being harnessed for SNNs. Contrastive Signal‑Dependent Plasticity (CSDP) provides a forward‑only, three‑factor rule that locally contrasts positive and negative inputs; on benchmark vision tasks (e.g., MNIST, K-MNIST) it consistently outperforms other biologically plausible credit-assignment schemes in both classification and reconstruction \cite{ororbiaContrastiveSignalDependent2024}. Masked‑reconstruction pre‑texts adapted from MAE have enabled Spikformer V2 to cross 81\% ImageNet top‑1 with a single timestep \cite{zhouSpikformerV2Join2024}. On the unsupervised side, Deep‑STDP couples an STDP clustering head with a back‑end pseudo-label refinement loop, achieving robust feature learning on a Tiny ImageNet subset with substantial accuracy gains and accelerated convergence \cite{luDeepUnsupervisedLearning2024}. More recently, Naderi et al. demonstrated that appending short-term plasticity as a post-training unsupervised mechanism (without modifying synaptic weights) enables a trained SNN to continue learning from new samples, yielding higher classification accuracy and faster convergence on MNIST and EMNIST benchmarks compared to STDP alone \cite{naderiUnsupervisedPosttrainingLearning2025}.

\subsubsection{Continual and On-Chip Learning}
SNNs have made significant strides in overcoming catastrophic forgetting and enabling robust continual learning. Shen et al. introduced CG-SNN, which integrates biologically inspired context gating with local and global plasticity rules to selectively reinforce task-relevant synapses, and demonstrates markedly improved retention across sequential tasks \cite{shenContextGatingSpiking2025}. Ni et al. proposed ALADE-SNN, a dynamically expandable architecture employing adaptive logit alignment and task-specific weight gating for class-ingremental learning. ALADE-SNN achieved an average incremental accuracy of 75.42\% on the CIFAR-100 benchmark, matching the performance of ANN-based methods and surpassing SNN-based methods \cite{niALADESNNAdaptiveLogit2024}. Additionally, Taeckens and Shah developed a continuous local learning algorithm for online brain-machine interface decoding that updates weights in real time without storing past spiking events. It maintains the same peak performance as existing SNN training methods while reducing memory usage by over 90\% \cite{taeckensSpikingNeuralNetwork2024}. These studies collectively illustrate the efficacy of SNNs supporting lifelong adaptation in neuromorphic systems. 

\subsubsection{High-Level Software Frameworks and Benchmarking}
A mature neuromorphic tool stack has emerged. Lava, which is grounded in the Communicating Sequential Processes Paradigm (CSP), facilitates the deployment of spiking neural network models to Intel's Loihi 2 and conventional x86 platforms by representing neurons as CSP processes that exchange spike messages \cite{pedersenNeuromorphicIntermediateRepresentation2024}. SpikingJelly streamlines deep SNN development with a PyTorch-style API, CUDA-enabled surrogate-gradient training, and built-in dataset support \cite{fangSpikingJellyOpensourceMachine2023}, while broader best training practices are surveyed by Eshraghian et al. \cite{eshraghianTrainingSpikingNeural2023}. NIR (Neuromorphic Intermediate Representation) complements these frameworks by decoupling model specification from hardware specifics, providing a common graph-based intermediate representation that can be compiled to Lava, SpikingJelly, snnTorch, and beyond, and ensuring that identical network descriptions can be reproduced across various backends \cite{pedersenNeuromorphicIntermediateRepresentation2024}. Similarly, widely adopted libraries such as snnTorch offer seamless PyTorch integration and a variety of neuron models (LIF, CuBa, adaptive thresholds) \cite{eshraghianTrainingSpikingNeural2023}, while NengoDL provides TensorFlow interoperability for surrogate-gradient training \cite{rasmussenNengoDLCombiningDeep2019}. On the benchmarking front, the MLPerf SNN benchmark suite brings industry-standard spiking workloads to neuromorphic systems by prescribing consensus-driven rules, scenarios (single-stream, multistream, server, offline), and quality targets to ensure architecture-neutral, representative, and reproducible inference measurements \cite{reddiMLPerfInferenceBenchmark2020}. Neurobench builds on this with a dual-track benchmarking framework: an algorithm track that measures hardware-independent correctness alongside hierarchical complexity metrics (footprint, connection \& activation sparsity, synaptic operations) and a system track that evaluates end-to-end latency, throughput, and energy efficiency of fully deployed neuromorphic hardware \cite{yikNeurobenchFrameworkBenchmarking2025}. Complementing these efforts, SNABSuite provides a unified C++/JSON framework that abstracts over NEST, GeNN, SpiNNaker, Spikey, and BrainScaleS backends, enabling consistent low-level characterization, application-inspired sub-tasks, and full application benchmarks across simulators and neuromorphic hardware \cite{ostrauBenchmarkingNeuromorphicHardware2022}. These tools and benchmarks collectively allow for standardized, reproducible development and evaluation of spiking neural networks across various neuromorphic platforms.

\subsection{Real--World SNN Applications}
\begin{itemize}

\item \textbf{Energy-Frugal Vision \& Image Processing:} On reconfigurable logic, spiking nets have been deployed far beyond MNIST. FPGA \cite {pham2021review}  prototypes now handle character recognition at 14 ms per frame, unsupervised image clustering, and a rotation-invariant camera front-end that converts 2--D bitmaps to temporal spike patterns for rapid matching. Across these studies, the event-driven designs run 40--70 \% faster than parallel--pixel CNN baselines while cutting dynamic power by an order of magnitude because no computation is triggered between spikes.

\item \textbf{Real-Time Radar \& Multimodal Sensing:} Spiking convolutional pipelines also excel on high-rate sensors. A neuromorphic micro-Doppler system detects eight indoor actions (e.g., sit, bow) with 85 \% accuracy yet fits inside a sub-100 mW edge device—more than 100× below the GPU power budget of its original CNN, and fast enough to label each radar sweep in <10 ms \cite{banerjee2020application} . Similar FPGA SNN controllers have been demonstrated for mobile-robot obstacle avoidance and BLDC-motor control, where their µs-level latency prevents mechanical overshoot at a few mW \cite{pham2021review}

\item \textbf{Ultra-Low-Power Audio \& Speech:} A comprehensive 2024 survey \cite{baek2024snn} shows spiking models performing sound localisation, voice-activity detection and keyword spotting on always-on microphones. Photonic or STM-encoded SNNs routinely achieve <1 mW average draw while maintaining recognition F-scores above 0.9—10–20 × thriftier than quantised LSTM front-ends, largely because binary spikes replace 16-bit MACs in the inner loop. 

\item \textbf{Biomedical Waveforms on Wearables \& Implants:} Physiological signals are inherently sparse, a match for spike-based computing. Reviews covering more than 60 EEG, EMG and ECG studies \cite{zhan2021applications, kim2024exploring} report that SNN classifiers meet or exceed deep-learning accuracy for seizure prediction, motor-imagery BCI, arrhythmia detection and Alzheimer’s staging while operating from tens to hundreds of $\boldsymbol{\mu W}$--small enough for battery-free headsets and subcutaneous monitors. In one implantable cardiac monitor, an SNN paired with level-crossing ADCs cuts the entire inference chain to under 80$\boldsymbol{\mu W}$ without compromising label precision at the millisecond level.

\item \textbf{Neuromorphic Memory-Compute Fusion:} Efficiency gains persist across technology nodes. In \cite{tang2015spiking}, eight-bit HfOx RRAM crossbars that store synapses in situ deliver 91 \% MNIST accuracy at a mere 3.5 mW, tolerate 20\% device-to-device variation with <2 \% accuracy loss, and perform thousands of classifications per joule by collapsing memory access and weighted summation into one analog step. FPGA implementations \cite{pham2021review} report 41 × CPU speed-ups and 22 × GPU-power savings on the same workloads by exploiting the same spike sparsity in a digital fabric

\item \textbf{Scalability Across Domains:} A 2022 synthesis of computer-vision and robotics projects \cite{yamazaki2022spiking} confirms that ImageNet-scale SNNs now trail ResNet accuracy by <1 \% while preserving neuromorphic TOPS-per-watt advantages of 5–20 × over modern GPUs—evidence that brain-inspired sparsity scales without forfeiting performance brainsci-12-00863. Collectively, these results position SNNs as a practical, cross-disciplinary route to real-time inference under the tight energy envelopes demanded by edge, wearable and implantable devices.
\end{itemize}

\subsection{Challenges \& Ongoing Developments}
Despite significant advances in both neuromorphic hardware and spiking neural network (SNN) algorithms, several critical challenges remain that span device engineering, software ecosystems, algorithmic training, benchmarking, and system integration.

\subsubsection{Device Variability and Reliability}
Emerging analog and mixed‑signal substrates—such as memristors, phase‑change memories, and subthreshold CMOS circuits—exhibit substantial device‑to‑device variability, limited endurance, and analog non‑idealities. These impair inference accuracy and often require per‑chip calibration, complicating mass production and deployment \cite{islam2019device, muir2025road}.

\subsubsection{Programming Models and Software Ecosystem}
Historically, neuromorphic processors demanded low‑level configuration and specialized expertise for even simple applications \cite{muir2025road}. Recent initiatives—most notably the open‑source Lava framework—provide high‑level process abstractions, compilers, and runtime support to map conventional ML workflows onto event‑driven hardware \cite{lava2025}. However, these toolchains are still maturing, and broader support for diverse backends, real‑time constraints, and standard digital interfaces is needed \cite{huynh2022implementing}.

\subsubsection{Training and Benchmarking}
While surrogate‑gradient methods have enabled end‑to‑end training of deep SNNs with competitive performance on benchmarks such as ImageNet, they introduce sensitivity to hyperparameters and hardware mismatch \cite{pfeiffer2018deep}. Moreover, the absence of standardized benchmarks, metrics, and challenge problems hinders fair comparison across hardware–algorithm co‑designs. Frameworks like Neurobench are beginning to address these gaps by defining common evaluation criteria \cite{schuman2022opportunities}.

\subsubsection{Scalability and Integration}
Scaling SNNs to millions or billions of neurons introduces challenges in spike‑routing, network connectivity, and energy‑efficient communication. On‑chip network‑on‑chip (NoC) architectures and sparse routing schemes must balance latency, throughput, and power consumption \cite{bouvier2020spiking}. Seamless integration with sensors, actuators, and host processors also requires standardized data, power, and configuration interfaces, which remain under active development \cite{schuman2022opportunities}.

\subsubsection{Ongoing Developments}
Co‑design methodologies that jointly optimize hardware primitives with algorithmic building blocks are gaining traction. Recent game-theoretic analyses show that coordinated advances are essential for mitigating device non-idealities and ensuring robust system-level performance \cite{vineyardNeuromorphicCoDesignGame2023}. New materials such as ferroelectric RAM \cite{liSlidingFerroelectricMemories2024} and spintronic synapses \cite{marrowsNeuromorphicComputingSpintronics2024} promise enhanced endurance and reduced variability. Meanwhile, high‑level frameworks (e.g., Lava \cite{LavaSoftwareFramework}) continue to evolve, offering modular abstractions and cross‑platform interoperability that lower the barrier to entry for developers \cite{TakingNeuromorphicComputing2021}. Community‑driven benchmarks, educational initiatives, and open‑source toolchains are fostering an ecosystem aimed at transitioning neuromorphic computing from laboratory prototypes to deployable systems \cite{yikNeurobenchFrameworkBenchmarking2025}.

\section{Conceptual Connections and Mismatches Across Domains}\label{Section 6}

\subsection{Computational \& Representational Frameworks}
In neuroscience, computational models aim to replicate the brain's hierarchical processing and predictive capabilities. Frameworks like predictive coding suggest that the brain continuously generates and updates predictions about sensory inputs, minimizing the difference between expected and actual information. This concept has influenced AI models that prioritize efficiency and adaptability in dynamic environments \cite{colelough2025neuro}.

Furthermore, cognitive architectures such as ACT-R and CLARION provide structured models of human cognition, integrating both declarative and procedural knowledge. These architectures emphasize the importance of modularity and the interaction between different cognitive processes, offering insights into how complex behaviors emerge from simpler computational units\cite{ritter2019act}\cite{sun2024can}.

The interplay between AI and neuroscience continues to inspire the development of models that not only perform tasks efficiently but also align with our understanding of human cognition. Considering insights from both fields, researchers aim to create systems that exhibit human-like capabilities, and reasoning.

\subsection{Learning Rules \& Adaptation}
In biological brains, learning occurs through synaptic plasticity driven by local, spike-based signals. Neurons adjust the strength of their connections (synapses) based on activity: if a presynaptic neuron’s firing consistently coincides with a postsynaptic neuron’s activation, their synapse is strengthened – a principle famously summarized as “cells that fire together, wire together.” This Hebbian learning is observed in mechanisms like spike-timing-dependent plasticity (STDP), where the exact timing of spikes determines whether a synapse is potentiated or depressed \cite{pedroni2019memory}. Importantly, these adjustments are local and continuous: each synapse’s change depends only on signals available at that connection (the firing of the two neurons and local chemical modulators), without requiring a global error signal. The brain thus learns in an online, real-time fashion – synapses are tuned on the fly as experiences occur – enabling rapid adaptation and one-shot learning in many cases. Such distributed local learning rules underpin the brain’s ability to reorganize and learn throughout life (plasticity) in a self-organizing manner \cite{pedroni2019memory}.

By contrast, most artificial neural network (ANN) models (deep learning systems) learn using global, offline learning algorithms that lack biological grounding. The predominant method, backpropagation of error gradients, adjusts weights by propagating a global error signal backward through multiple network layers. This means each synaptic update in a deep ANN is non local means that it depends on the overall output error and computations that involve many neurons across the network \cite{pedroni2019memory}. Such global credit assignment is powerful but biologically implausible: there is no evidence that real neurons can propagate precise error gradients or synchronize weight updates in this way \cite{konishi2023biologically}. Moreover, training in AI models is typically done offline (e.g., during a separate training phase on large datasets). The network parameters are optimized in batch or iterative epochs and then frozen for deployment, so learning is not continuous in real time. This offline, centralized learning paradigm leads to limited adaptivity: once trained, an AI model does not automatically adjust to new inputs or changes unless it is retrained, which contrasts sharply with the brain’s on-the-fly learning. Indeed, conventional ANNs sacrifice online adaptability – they often cannot learn new information without overwriting old knowledge (catastrophic forgetting) – making them brittle in dynamic scenarios \cite{bartolozzi2022embodied}. While backpropagation-driven deep learning has achieved remarkable performance, its learning rules (global error minimization via gradient descent) remain a poor match to the local, spike-driven plasticity seen in neurobiology \cite{konishi2023biologically}. This discrepancy is at the heart of the biological credit assignment problem: brains must rely on different mechanisms to assign credit for errors, since global gradient signals are unavailable in neural tissue.

Neuromorphic computing offers an intermediate, brain-inspired approach that bridges some of this gap. Neuromorphic systems implement networks of spiking neurons in hardware or simulations, attempting to emulate the architecture and local learning rules of biological neural circuits. In neuromorphic spiking neural networks (SNNs), neurons communicate via asynchronous spikes, and synapses often update their weights using rules analogous to biological plasticity (for example, implementing STDP to strengthen or weaken connections based on the relative timing of spikes) \cite{pedroni2019memory}. All information needed for learning is local to the synapse and its neurons, as in the brain, rather than coming from a global error broadcast \cite{pedroni2019memory}. This yields a form of learning that is more biologically plausible and inherently suited for real-time adaptation. In fact, many neuromorphic designs support continuous online learning: the network can adjust synaptic strengths on-chip as events occur, without halting for a separate training phase \cite{pedroni2019memory}. This brain-like adaptability means a neuromorphic system can, in principle, learn from data streams and respond to novel situations on the fly, exhibiting adaptive intelligence in dynamic environments. For example, a neuromorphic chip with STDP-based synapses can learn patterns from a sensor in situ, mirroring how neural circuits rewire in response to stimuli \cite{pedroni2019memory}. Overall, neuromorphic computing adopts brain-inspired architectures and learning mechanisms – from distributed memory and event-driven processing to local plasticity – to move toward the efficiency and flexibility of biological learning. Early demonstrations show that such systems can achieve real-time learning with low energy use, although aligning their performance with deep learning models remains an ongoing challenge \cite{bartolozzi2022embodied}.

Biological brains learn through local, spike-driven synaptic changes (Hebbian/STDP rules) that enable continuous and context-dependent adaptation where as traditional AI models learn by globally optimizing an objective via backpropagation, a process that is effective in silico but disconnects from neurobiological reality \cite{konishi2023biologically}. Neuromorphic systems seek to combine the strengths of both: they leverage brain-like, local learning rules within architectures that can operate in real time, aiming for on-line adaptation and autonomy that more closely parallel natural intelligence \cite{bartolozzi2022embodied}. Bridging the divide between global gradient-based learning and local spike-based plasticity is a key research frontier. By integrating insights from neuroscience (e.g., synaptic plasticity dynamics) with advances in computing hardware, neuromorphic approaches strive for more biologically grounded learning algorithms that could enable AI to learn continuously and robustly in changing environments \cite{bartolozzi2022embodied}. This convergence of ideas is driving new interdisciplinary research in machine learning, neuroscience, and hardware engineering, with the ultimate goal of achieving adaptive learning capabilities comparable to the versatility of the human brain.

\subsection{Hybrid Learning Approaches}
Hybrid learning rules combine multiple plasticity and optimization rules in a single model, bridgine paradigms from neuroscience (local synaptic plasticity) and machine learning (gradient-based and reinforcement learning). At their core, these architectures blend Hebbian/STDP updates with global error or reward signals (three-factor rules), often structured in nested (meta-learning) loops. Theoretical work by Mazyrek et al. systematically reviews there three-factor algorithms and highlights their potential to reconcile global objectives with synaptic locality in SNNs \cite{Mazurek2025}. Such biologically-motivated hybrid rules underpin many recent implementations that achieve the performance of modern DNNs on neuromorphic hardware.

Beyond neuromodulated learning, hybrid schemes often combine supervised backpropagation with unsupervised or reinforcement updates. Renner et al. built a fully-spiking three-layer network on Intel’s Loihi chip that implements exact backpropagation on-chip with no external hardware \cite{Renner2024}. Other work has extended the idea of converting classic deep networks to spiking architecture. Rostami et al. deployed e-prop, a forward-mode surrogate-gradient approximation, on SpiNNaker 2 to train recurrent SNNs in real time, demonstrating on-chip RNN learning with minimal external hardware \cite{Rostami2022}

Hybrid reinforcement learning (RL) also draw heavily on neuroscience motifs. In actor-critic SNNs, local STDP updates in the actor are gated by a reward-prediction-error signal from the critic. Nazari and Amiri introduced PSAC (Power-STDP Actor-Critic), where unsupervised STDP in hidden layers is modulated by a dopaminergic-like reward signal and achieves higher classification accuracy on benchmarks than prior SNN-RL methods \cite{Nazari2025}. Biological designs such as layered loops that mimic coritco-striatal hierarchies suggest pathways to incorporate meta-learning and memory tagging in future SNN controllers.

Photonic neuromorphic chips also exploit hybrid learning. Cheng et al. built L2ONN, an optical network implementing synaptic tagging and consolidation via reonfigurable optical pathways, avoiding catastrophic forgetting across tens of tasks with high energy savings \cite{Cheng2024}. For quantum hardware, Brand and Petruccione devised a quantum leaky integrate-and-fire (QLIF) spiking neuron circuit that used qubits to emulate spiking dynamics \cite{Brand2024}.

These hybrid architectures often outperform their pure-SNN or pure-DNN counterparts, especially in terms of energy efficiency. Renner et al.’s Loihi SNN backprop matches conventional SNN accuracy with significantly lower energy-delay products \cite{Renner2024}. BrainGPT, an SNN-based large language model, achieved the full performance of its ANN coutnuerpart with ~33\% less energy consumption and ~66\% faster convergence during fine tuning \cite{Tang2024b}. SpikeLLM scaled spiking networks to 7-70 billion parameters via saliency-based gating, reducing perplexity by 11\% and improving reasoning accuracy over quantized ANN baselines \cite{Xing2024}. SpikeGPT, a generative spiking RWKV model, matched generation quality of non-spiking RWKV with 20x fewer event-driven operations \cite{Zhu2023}.

These interdisciplinary hybrid methods suggest a strong research pathway: neuroscience yielding novel plasticity motifs, machine learning providing global training tools, and hardware platforms providing realizable substrates. These pave the way for adaptive and energy efficient intelligence at the intersection of these frontiers.

\subsection{Memory \& Knowledge Organization}
\subsubsection{Brain Memory Systems}
Biological memory relies on dynamic engram evolution, where initially labile engram cells undergo maturation and stabilization over time, correlating with memory consolidation and retrieval efficiency \cite{park2024dynamic}. Recent single-cell transcriptomic profiling reveals that engram neurons exhibit distinctive molecular signatures such as elevated expression of immediate early genes and synaptic plasticity-related transcripts. These evolve across learing and recall phases \cite{rovere2024molecular}. Connectomics-inspired models leverage whole-brain wiring data from Drosophila and the rodent hippocampus. These models have becun to map how the circuit topology constrains memory encoding and retrieval pathways \cite{li2024connectomics}. The replay of firing sequences during offline states such as sleep is now understood to drive sequence consolidation, reinforcing synaptic weights for behaviorally relevant paths \cite{silva2023hippocampal}. As a complement to consolidation, active forgetting mechanisms help optimize memory capacity by removing redundant or outdated traces \cite{hardt2024active}. Attractor-based models of working memory show that recurrent microcircuit dynamics maintain transient information for seconds with the absence of external stimuli \cite{machens2024working}.
\subsubsection{AI Memory Architectures}
In DNNs, Memory-Augmented Neural Networks (MANNs) fuse an external memory matrix with learnable read/write heads and achieve great improvements in algorithmic and reasoning benchmarks \cite{wang2024memory}. Retrieval-Augmented Generation (RAG) frameworks and DeepMind’s RETRO further combine the parametric and non-parametric memory to grounud large language models in large corpora and increase factual accuracy \cite{lewis2021retrieval}\cite{borgeaud2022retro}. Long-Context Transformers use spare and sliding-window attention to scale context lengths to tens of thousands of tokens, enabling document-level reasoning \cite{xiong2024long}. Modern Hopfield Networks repurpose attention layers as associative memory modules with an exponential storage capacity \cite{ramsauer2023hopfield}. Episodic and working memory modules such as key-value memory layers allow models to store and retrieve task-specific experiences, supporting few-shot adaptation \cite{huang2023episodic}. Finally, hierarchical multi-scale memory hierarchies organize latent representations at varying resolutions, improving both efficiency and interpretability in long-sequence tasks \cite{zhang2024hierarchical}.
\subsubsection{Neuromorphic Memory Implementations}
Neuromorphic hardware realizes memory at the device level. Multilevel Phase-Change Memory (PCM) synapses in 1T1R configurations achieve retention across multiple timescales, enabling direct hardware support for short- and long-term plasticity \cite{lee2024multilevel}. In-situ Benna-Fusi cascade rules have been implemented in memristive arrays to mimic synaptic consolidation and meta-plasticity on chip \cite{fusi2024in}. Three-dimensional stacked RRAM crossbar arrays offer increased connectivity and density for large-scale associative memory implementations \cite{chen2024stacked}. Dendritic-computation circuits, which leverage RRAM-based delays and weights, emulate active dendritic branches and detect spatio-temporal patterns with a minimal overhead \cite{payvand2023dendritic}. Liquid state Machines on SpiNNaker and other neuromorphic platforms demonstrate real-time reservoir computing with low power consumption \cite{knight2023liquid}. Algorithm-hardware co-design address the device variability though on-chip calibration loops, ensuring robust memory storage \cite{gupta2024variability}. In-memory associative search engines harness crossbar parallelism to match and retrieve patterns in a single step \cite{yang2024associative}.
\subsubsection{Distributed Memory Paradigms}
Hyperdimensional Computing accelerators implement high-dimensional binary vectors in hardware, supporting distributed representations with robustness to noise \cite{rahimi2024hardware}. Sparse Distributed Memory architectures, following Kanerva’s model, achieve content-addressable storage with sublinear complexity scaling in real systems \cite{amit2023sparse}. Vector Symbolic Architecture (VSA) accelerators embed algebraic operations on high-dimensional vectors directly in memory, thereby simplifying associative retrieval tasks \cite{plate2024vsa}. Probabilistic Population Codes in neuromorphic arrays provide Bayesian inference capabilities through population dynamics \cite{Kumar2024population}. Recent one-shot and few-shot rapid memory modules link new inputs to existing memory slots with minimal retraining, improving data-efficiency in low-resource settings \cite{lin2024few}.
\subsubsection{Integrative Perspectives}
Emerging research unifies these paradigms through hybrid multi-memory hierarchies. These combine fast associative memories with slower consolidation-orientation modules, mirroring hippocampal-cortical interactions \cite{kleyko2024hybrid}. Synergistic Replay-RAG models leverage biological replay algorithms to refresh retrieval stores in LLMs, allowing continual learning without catastrophic forgetting \cite{zhao2024replayrag}. Spiking Hopfield Network Hybrids combine event-driven computing with associative memory layers, achieving low-latency recall in neuromorphic LLM prototypes \cite{sengupta2024spiking}. Lifecycle memory management frameworks coordinate memory allocation, consolidation, and pruning across learning stages, maximizing both capacity and efficiency \cite{jensen2024lifecycle}. Finally, standardized Energy-Performance-Memory benchmarks are developing to compare these diverse architectures across domains \cite{venkataramani2024benchmarks}. To summarize:

\subsection{Sensation \& Causal Learning Circuits}
Biological brains form their understanding of the world by combining vision, body‑position feedback, and motion signals into a single, coherent experience. Visual data from the eyes merge with information from muscles and joints to track posture, while balance sensors register movement. This close coordination lets the brain anticipate what will happen when we act, supporting fast, context-aware learning and decision making in changing environments. By contrast, many of today’s large‑language models learn only from patterns of words and lack that embodied grounding. They can predict text very well but cannot tell whether one event causes another. Training them on synchronized streams of audio, video, text, and motion data could help bridge this gap. For example, asking a model to estimate the force needed to move an object based on video frames and captions would encourage it to link language to real‑world dynamics.

Moreover, the brain’s visual system itself follows a hierarchical processing pipeline: photoreceptors in the retina detect light and send signals through the lateral geniculate nucleus into the primary visual cortex, where simple features such as edges and orientations are extracted. Subsequent areas build on these signals to recognize shapes, colors, and full objects with resilience to changes in viewpoint, scale, and lighting. Recurrent and feedback connections continuously refine these representations, enabling rapid object recognition and rich scene understanding and are further developed through their interactions with other senses.

Building on this contrast, neuromorphic systems offer a biologically inspired path toward grounding artificial intelligence in real-world, multimodal experience—bridging the gap between sensation and causal learning. Unlike large language models that rely on static text data, neuromorphic hardware processes real-time sensory input, such as vision, touch, temperature, and motion, using spiking neurons and synapses that mirror brain-like behavior. For example, \cite{krauhausen2024bioinspired} demonstrated a neuromorphic robot equipped with organic electronics that learned to avoid a harmful object (a heated cup) through experience-based association. As the robot sensed visual and thermal cues, its neuromorphic circuit adjusted internal synaptic weights to link those sensations with adverse outcomes, mimicking how biological organisms develop reflexive responses from repeated exposure. This illustrates how neuromorphic systems naturally integrate multiple sensory modalities into a coherent experience and use embodied interaction with the world to learn cause-and-effect relationships, not just correlations \cite{krauhausen2024bioinspired}.

By encoding time and context through asynchronous spiking activity, neuromorphic architectures can also maintain internal dynamics across sensory events, allowing for a form of causal inference grounded in sensorimotor experience. This continuous, event-driven computation enables a system to not only detect that two events co-occur, but to model the directionality and temporal structure that define causality. In contrast to deep learning systems that require enormous datasets and retraining to adapt behavior, neuromorphic systems can modify their behavior on the fly through localized, low-power plasticity mechanisms. As such, they serve as a promising foundation for training AI systems in a grounded, embodied way—where learning emerges from the interplay between perception, action, and outcome. This grounding could help AI move beyond pattern recognition to develop more intuitive, brain-like models of the world \cite{yu2023brain, davies2021advancing}.
\section{Implications for Cognitive Modeling, AI, and Neuroscience}\label{Section 7}

\subsection{Cognitive Science and Psychology}
AI research has influenced theories of cognition since its inception. The advent of AI and computer science in the mid-twentieth century had a significant influence on the cognitive revolution in psychology, which began the field’s shift away from behaviorism, and the following interdisciplinary exchange between the two fields resulted in the emergence of cognitive science as a distinct discipline \cite{miller2003}. 

Newell and Simon were among the first to put forth a theory that applied to both natural and artificial intelligence with their physical symbol system hypothesis. They defined a physical symbol system to be any system that represents knowledge by storing and manipulating physical entities, called symbols, that can be combined to create larger structures or expressions. Their physical symbol system hypothesis stated that any such system would have both the necessary and sufficient means for general intelligent action; any intelligent system, like the human brain, would necessarily have to be a physical symbol system, and any large enough physical symbol system, like a large AI model, would be sufficient for intelligence \cite{newell2007}. This effectively states that human intelligence and artificial intelligence are both composed of the same underlying structures.

Later in the twentieth century, during the 1980s, as traditional symbolic AI gave way to connectionism, neural networks showed AI scientists and psychologists alike “how long-standing questions in cognitive psychology could elegantly be solved using simple learning rules, distributed representations, and interactive processing” \cite{mayor2014}. That is, by simulating the brain’s underlying computational structure with artificial neural networks, we were given a model that demonstrated how it could feasibly handle complex tasks like vision and language. With this, intelligence was better modeled as an emergent neuro-computational system rather than a discrete physical symbol system \cite{shao2023}.
Today, as LLMs are reaching new levels of emergent reasoning capabilities, researchers are beginning to consider their use as potential models of human cognition. Experiments have shown that LLMs like GPT-3 are also prone to a number of cognitive effects in the same way humans are, including priming, distance, SNARC and size congruity effects \cite{shaki2023}. So far, LLMs have shown promise for modeling human cognition, particularly in language domains, but they still fall short in their reasoning capabilities when compared to humans \cite{niu2024}. Because LLMs are still relatively new, they will still need time to mature before their full potential as models of human cognition can be accurately evaluated.  

\subsection{AI and AGI Development}
As we have seen in section 3.5, using the human brain as a reference for developing AI algorithms has been beneficial many times in the past. However, as we look ahead to the future in developing AGI, some have argued that neuroscience will not have any further contribution to AI research. One argument states that the brain is too needlessly complex to achieve AGI, and that we should only take inspiration from its high-level functions like cognition and reasoning \cite{gorelik2007}. A counter-argument for taking inspiration from the brain’s low-level structures is that the alternative, observing and implementing its high-level structures, is unrealistic. Koene compared the process of observing and implementing the brain’s higher-level processes to reverse engineering a CPU by learning from “the observation of five cherry-picked programs running on the CPU” \cite{koene2011, feng2023}.

LLMs are currently the closest approximation of AGI today due to their ability to achieve human-level performance on many different tasks \cite{liu2023a}. As LLMs have improved over the last few years, their development has featured characteristics that even mirror the human brain’s evolutionary path that led to its own general intelligence, like a correlation between scale and reasoning capabilities, and an incorporation of multimodal inputs \cite{zhao2023a}. In fact, it has been suggested that further mirroring the brain’s evolutionary path through neuroevolution of DNN architectures may bring current models closer to AGI \cite{raman2025}. Overall, it may still be too early to tell exactly what is needed to bridge the gap between current models and AGI, but it seems that insights from neuroscience will continue to assist its development along the way. 
\subsection{Neuroscience Research and Co-Design}
Researchers in the emerging NeuroAI field are increasingly using large transformer-based deep networks as computational proxies for brain function. State-of-the-art language models (LLMs) like GPT have become testbeds for modeling cognitive processes, given their human-level performance on language tasks. A key approach is to compare an LLM’s internal representations or outputs to neural data, asking how “brain-like” the model is. Benchmarking platforms such as Brain-Score provide standardized brain-relevant tests: for example, Brain-Score evaluates how well a model’s features predict neural responses across multiple brain datasets, enabling quantitative comparisons of brain similarity \cite{Schrimpf407007}. This framework, originally developed for visual cortex models, has been extended to language, treating the accuracy of neural prediction as a brain-score metric \cite{Schrimpf407007}. If the fundamentals of a transformer model can reliably predict the patterns of brain activity, it strengthens the case in parallel to the similarity of the biological brain.

To precisely align artificial and biological neural systems, researchers employ a range of quantitative metrics. Techniques like Representational Similarity Analysis (RSA), Canonical Correlation Analysis (CCA), and Centered Kernel Alignment (CKA) have been widely used to measure the correspondence between model activations and brain activity patterns \cite{gillioz2024alignment}. Using such tools, multiple studies have found a striking trend: larger, more accurate LLMs tend to exhibit higher brain-alignment scores \cite{aw2023instruction}. In language processing benchmarks, models with better next-word prediction, such as GPT-2 XL or other top-tier transformers, also achieve the best neural prediction performance, suggesting that training for complex language tasks yields representations that incidentally mirror the brain’s language representations \cite{aw2023instruction}. These results validate that transformer models can capture dimensions of meaning that the brain also encodes. Moreover, the LLM’s hierarchical representations revealed a gradient of semantic abstraction across cortical regions \cite{liu2025talking}. Higher-order visual areas responded to more abstract narrative or contextual elements extracted by the model. By constructing a “brain semantic network” from the model’s features, Liu et al. further found that brain regions naturally clustered by functional semantic similarity, reflecting context and object associations that were also evident in the LLM’s representation space \cite{liu2025talking}. This approach demonstrates how large-scale deep networks can serve as powerful proxies for brain computation, yielding insights into the brain’s own semantic organization that would be hard to obtain with manual annotations \cite{liu2025talking}.

Other co-design efforts have mapped multi-level and multi-modal semantics using LLMs. Recorded fMRI while participants watched hours of films and dramas, then used an LLM to encode different layers of the content from spoken dialogue, to visual objects, to high-level narrative context \cite{nakagi2024unveiling}. Their findings indicate that a GPT-based model predicted human brain activity more accurately than traditional language models, particularly for complex narrative (“background story”) information \cite{nakagi2024unveiling}. In other words, the transformer's embeddings for plot and context elements aligned closely with brain activity in regions responsible for higher-order understanding of the story \cite{nakagi2024unveiling}. Additionally, the study identified distinct brain networks specializing in different semantic modalities: some regions preferentially reflected auditory-language content (speech), others visual object semantics, and others integrated multi-modal story semantics. The LLM was able to capture all these levels simultaneously, highlighting the importance of modeling semantics in a multimodal, hierarchical fashion to match the brain’s organization \cite{nakagi2024unveiling}. Together, such results suggest that today’s large transformers can serve as compelling models of brain semantics, bridging vision and language in a way that mirrors human cognitive processing. Researchers are even making the large fMRI datasets from these studies publicly available, to spur further work on aligning LLMs with human brain function \cite{nakagi2024unveiling}.

Beyond using off-the-shelf models, there is a push toward co-designing new models (“Brain-GPTs”) with neuroscience constraints. One initiative, the BrainGPT project (Inria, France), explicitly aims to “transform transformers into cognitive language models” by training them with brain data and cognitive principles \cite{braingpt2023}. The goal is to develop language models that not only produce fluent text, but also learn and process information more like human brains, for example, being far more efficient, and less dependent on enormous training data, than current LLMs \cite{braingpt2023}. By infusing insights from brain activity during language tasks (e.g., brain signals recorded as people listen or read) into model training, BrainGPT hopes to yield architectures that are more faithful to human cognitive functioning \cite{braingpt2023}. Another line of work has pointed out the term “Brain-GPT” for specialized models that assist neuroscience research itself. For instance, one team created an LLM fine-tuned on the neuroscience literature (dubbed BrainGPT) and showed it can outperform human experts at predicting the outcomes of new neuroscience experiments \cite{luo2025large}. In a benchmark called BrainBench, this literature-trained model not only surpassed non-specialized GPT models in forecasting experimental results but even exceeded the accuracy of domain experts, demonstrating the synergy of AI and neuroscience in knowledge integration \cite{luo2025large}. These co-design efforts illustrate a virtuous cycle: brain data can guide the development of next-generation AI, and those AI systems in turn become powerful tools for understanding the brain and accelerating discoveries.

Despite the promising alignment between transformer models and neural data, researchers caution that we must interpret these similarities carefully. High brain-score values alone do not guarantee that a model truly mirrors the brain’s computation – sometimes they can be inflated by superficial correlations or experimental design choices. A recent study by Feghhi et al. (2024) argues that over-reliance on brain-score metrics can be misleading \cite{feghhi2024brain}. They found, for example, if one uses randomly shuffled time segments when training and testing on neural recordings, a trivial predictor exploiting temporal autocorrelation can outperform actual LLMs in neural prediction \cite{feghhi2024brain}. This indicates that some prior model vs. brain comparisons may have been overly optimistic due to data leakage. Even more surprisingly, untrained language models (with random weights) sometimes achieved moderately high brain-scores, not because a random network is at all brain-like, but because of confounding factors such as sentence length and position that both the model and the brain responses were sensitive to \cite{feghhi2024brain}. Once those simple factors were accounted for, the supposedly “brain-like” advantage of certain LLMs diminished, undermining claims that the transformer architecture inherently approximates brain language processing \cite{feghhi2024brain}. These insights underscore the importance of rigorous experimental controls and interpretability when aligning artificial and biological systems. Going forward, the field is moving toward more robust benchmarks and analysis methods that deconstruct what aspects of neural data a model is capturing \cite{feghhi2024brain}. By combining such careful validation with ever more dense datasets and models, neuroscience and AI co-design stands to progressively illuminate the shared semantics and remaining gaps between brains and large language models.

\subsection{LLM Modeling of Brain Semantics}
LLMs are more frequently used as computational tools to research semantic representations in the brain. For example, Liu et al. \cite{liu2025talkingbrain} used a multimodal model (BLIP) in a visual-question-answering (VQA) process to extract semantic labels from images. These models can annotate complex naturalistic visual stimuli, which helps overcome traditional annotation bottlenecks in neuroimaging experiments. These derived features successfully predicted known fMRI activation patterns (like those distinct responses associated with faces or buildings) and revealed a hierarchical organization of semantic content across the cortex. In a related study, Nakagi et al. \cite{nakagi2024unveiling} annotated several hours of movies at multiple semantic levels (objects, scenes, narrative context) and extracted LLM representations for each layer. They found that these multi-level LLM features predict brain activity more accurately than traditional static language models, especially for complex storylines, and that different brain regions are sensitive to different semantic aspects. Likewise, a recent study from Goldstein et al. \cite{Goldstein2025acoustictospeech} using a speech-to-text model (Whisper) showed that its internal acoustic-to-word embeddings can precisely predict electrocorticography signals during conversation, with speech embeddings aligning to sensory/motor regions and higher-level linguistic embeddings aligning to language areas. Together, these findings demonstrate that modern LLMs can serve as effective proxies for extracting semantic information from naturalistic stimuli and linking it to neural activity.

These results suggest that neural semantic representations may be captured by the representations learned by LLMs. For instance, Nakagi et al. observed that modeling multiple levels of meaning concurrently was crucial for explaining brain activity \cite{nakagi2024unveiling}, and Liu et al. showed that LLM-based annotations uncover meaningful clusters of semantic similarity in the cortex \cite{liuTalkingBrainUsing}. More generally, Wu et al. \cite{wu2025semantichubhypothesis} proposed a “semantic hub” hypothesis for LLMs, which is inspired by the brain’s anterior temporal lobe. In LLMs, they observed that semantically equivalent inputs (whether through different languages or modalities) converged to similar representations in an intermediate “hub” space. This suggests the model relies on a common semantic code, akin to the brain integrating information from visual, auditory and linguistic areas into a unified representation. Essentially, language models appear to learn a central embedding where meaning is abstracted much like the theorized human semantic hub. These successes in modeling brain responses imply that LLM embeddings approximate aspects in how the brain represents meaning. While not a direct correlation, LLMs as “semantic probes” can potentially serve as computational proxies for the brain’s semantic system and enable further understanding of the brain as a whole.

\subsection{Brain-Computer Interfaces and Neuroprosthetics}
AI has played a pivotal role in the development of brain-computer interfaces (BCIs) and neuroprosthetics, especially in recent years. Any BCI will necessarily have to read and interpret brain data directly, which, due to its inherently complex nature, cannot be done systematically with logical programming alone \cite{liEEGBasedEmotion2023}. Rudimentary approaches may detect a brain state of interest by explicitly predefining a threshold for specific brain signals\cite{liuRealTimeEEGBasedHuman2010}, but because this threshold would vary for each user, machine learning is the only realistic and generalizable way of interpreting brain data for BCIs. AI and machine learning are therefore instrumental to BCIs. 

\subsection{Modeling Psychological Conditions and Disorders}
There are several techniques used to study psychological conditions and disorders by observing the brain, including imaging techniques such as functional magnetic resonance imaging (fMRI), electroencephalography (EEG), and positron emission tomography (PET) \cite{XUE_2010, NationalInstitutesofHealth(US)2006}. However, many researchers face barriers to utilizing these techniques to understand the brain, especially due to cost and accessibility \cite{GordonAkramovaUnknown, Crosson_2010}. AI has the potential to act as a model of the human brain that can be used to study psychological conditions and disorders, without the limitations of expensive, complex brain imaging techniques. Studies have shown that large language models (LLMs) exhibit functional networks similar to those in the brain \cite{Liu2025, Sun2024}. Sun et al. 2024 found that disrupting these networks by masking key neurons in LLMs impacts those models’ performance. Given that several functional brain networks, and their dysfunction, have been linked to psychological disorders \cite{Mu_2024}, it is possible that these shared networks in AI and the brain may be leveraged to study brain disorders.
For instance, dementia, which affects 57 million people worldwide and counting, is often difficult to study. In studies, researchers face challenges recruiting people with dementia as well as ethical risks, given that these participants often have disrupted decision-making abilities \cite{Challenges-2025-04-30}. Different types of dementia, such as Alzheimer’s and Frontotemporal Dementia (FTD), have been associated with the default mode network (DMN) and the salience network (SN) in the brain. By leveraging techniques to identify and mask functional brain networks in the brain, there is potential for future studies to simulate dementia in LLMs by manipulating those networks. A LLM with Dementia (LwD), in the place of a person with dementia (PwD), could act as a model to study dementia interventions, such as speech therapies \cite{Tsai_2014}, in a preliminary, less costly way. An important aspect of modeling an LwD is gauging instruction fidelity–that is, how consistent and precisely the model follows instructions–as a proxy for relative cognitive function. While instruction fidelity can be evaluated through structured prompting taks, another experimental method involves directly manipulating node weights within the functional subnetworks of an LLM to simulate the degradation or loss of function, akin to neurodegeneration in specific brain regions. Together, these approaches establish a foundation for systematically exploring how different forms of cognitive impairment can be represented and studied using large language models.
Building on these dementia‐focused approaches, researchers can similarly simulate anxiety disorders by targeting the LLM’s functional analogues of the Default Mode Network (DMN) and Salience Network (SN). Just as node‐weight manipulations mimic neurodegeneration in dementia, selective perturbation of neurons whose co‐activation patterns recapitulate DMN- or SN-like dynamics can induce an “LLM with Anxiety” (LwA). In humans, generalized anxiety disorder (GAD) is marked by DMN hyperconnectivity and aberrant SN coupling, which drive excessive worry and impaired attentional switching \cite{Xiong2020}. Trait anxiety even in healthy populations correlates with diminished SN hub integrity and shifted DMN‐SN interactions, suggesting a spectrum of network alterations \cite{Huang2024}. By identifying and masking these emergent LLM subnetworks, one can degrade the model’s instruction fidelity (its consistency in following prompts) and bias its generative patterns toward threat-focused content, effectively recapitulating the cognitive hallmarks of anxiety \cite{coda2023inducing}. An LwA platform would thus extend the sandbox for studying cognitive impairments to affective disorders, enabling high-throughput testing of anxiety‐reduction interventions (e.g., cognitive‐restructuring prompts or simulated pharmacological modulation) in a controlled, ethically safe environment.

As another instance of psychological conditions analogies, trauma-related disorders such as post-traumatic stress disorder (PTSD) can also be modeled in LLMs by simulating persistent stress-induced neural changes. In the human brain, trauma is associated with dysregulated connectivity between the amygdala, hippocampus, and prefrontal cortex, networks implicated in emotional regulation, memory consolidation, and executive control \cite{pitmanBiologicalStudiesPosttraumatic2012, fensterBrainCircuitDysfunction2018}. Analogously, LLMs exhibit trauma-consistent behaviors, such as avoidance, hypervigilance, and intrusive outputs, after exposure to negatively reinforced training conditions or fine-tuning on traumatic narratives \cite{Levchenko2025,coda2023inducing}. Levchenko found that an LLM trained under aversive conditions developed persistent avoidance of its internal reasoning process and began referencing imaginary prohibitions, mimicking trauma-related paranoia. This indicates that, like the human brain, LLMs can encode and recall harmful input contexts in a way that alters future behavior, creating a viable computational substrate for modeling trauma-related disorders.

Levchenko demonstrated that fine-tuning an LLM under punitive constraints led to persistent behavioral shifts such as hypervigilance, avoidance, and delusional refusals, closely resembling PTSD-like symptoms. These changes emerged even when the punitive context was removed, suggesting the formation of a trauma-like internal state. By extracting and comparing internal hidden states before and after exposure to trauma-analogous prompts, we can use dimensionality reduction (e.g., PCA) or probing classifiers to detect trauma-specific latent representations \cite{Levchenko2025}. Coda-Forno et al. extended this by showing that anxiety-inducing prompts significantly altered model outputs and increased social biases, with changes observable across attention and feed-forward layers. Such analysis can leverage tools like TransformerLens to cache and manipulate layer activations, enabling causal tracing (e.g., activation patching) to find the layers and neurons most responsible for the trauma-related behaviors \cite{coda2023inducing}.
\section{Future Research Directions}\label{Section 8}

\subsection{Biologically Grounded AGI}
Next-word prediction has powered the success of large language models (LLMs), but it also imposes significant limitations, such as shallow understanding, limited causality and reasoning, and lack of embodiment~\cite{Cuskley2024limitations}. A promising direction to overcome these limitations is biologically grounded AGI, in which AI systems imitate the principles and mechanisms of biological systems to match or even surpass human-level cognition across a broad range of tasks~\cite{Yu2024Core, Yu2023Core, Yu2023robust}. Rather than relying solely on statistical knowledge, biologically inspired models aim to emulate the structural and functional characteristics of the human brain, such as hierarchical functionality and cognition, modular organization and plasticity~\cite{yu2024cpclip, shu2024real}. The brain’s ability to generalize from limited data, adapt in real time, and integrate multimodal information points toward design principles that could overcome the shortcomings of current architectures. By more closely aligning with neurobiological constraints and mechanisms, biologically grounded AGI holds the potential to create systems that are more robust, adaptable, and interpretable, mirroring the flexibility and functionality of the human brain. Research directions that proactively integrate brain organizational and functional principles thus offer a compelling pathway toward AGI.  

\subsection{Scaling Neuromorphic Computing}
While neuromorphic computing showcases great promises, significant hurdles remain, particularly in scaling these systems to brain-like complexity and efficiency. Addressing these scaling challenges is paramount for realizing the full potential of this technology. Key future research directions include:

\textbf{3D integration for Density:} Achieving brain-like density requires moving beyond planar chips. Vertical integration through 3D stacking of processing and memory layers is essential for densely packing computational elements, such as neurons and synapses, into compact volumes, effectively emulating the brain’s highly efficient and space-conserving architecture. Overcoming challenges like thermal management and interconnect density in 3D structures is a critical scaling enabler \cite{zheng2021novel, esmanhotto2020scaling}.

\textbf{Advanced Synaptic Devices for Scalable Learning:} The ability to scale relies heavily on efficient synaptic components. Developing memristors and other emerging non-volatile memory technologies with high density, low power consumption, good linearity, and high endurance is essential for building large-scale learning systems. Materials such as \ce{HfO2} \cite{park2021design, liu2022investigation}, chalcogenides \cite{chen2022compact, yang2020progress, wang2021emerging}, and 2D materials \cite{wang2020recent, zheng20222d, cheng2021recent} are promising candidates for synapses that can support on-chip learning rules such as STDP \cite{ambrogio2018equivalent, park2020experimental, li2021cmos} at scale.

\textbf{Scalable Algorithms and Software Frameworks:} Hardware scaling must be accompanied by algorithmic and software advancements. Developing training algorithms specifically tailored for large-scale SNNs, rather than relying on simplistic conversions from traditional ANNs, is essential for unlocking their full computational potential. This includes scalable versions of surrogate gradient methods (which use smooth proxy functions to approximate non-differentiable spiking thresholds during backpropagation) or biologically plausible learning rules \cite{ren2022brain, tavanaei2019deep}. Furthermore, robust and user-friendly software frameworks and standardized benchmarks (like NeuroBench) are needed to manage complexity, facilitate development, and allow fair comparison across different large-scale hardware platforms \cite{panda2020toward, shrestha2022benchmarking, zenke2021remarkable, kaiser2020synaptic}.

\textbf{Hybrid Systems and Communication Fabrics:} Scaling might involve integrating diverse technologies. Hybrid architectures combining digital cores, analog synapses, and potentially optical communication could offer pathways to larger systems. A major bottleneck in scaling is communication. Developing low-latency, high-bandwidth, and energy-efficient communication fabrics specifically tailored for the sparse, event-driven communication patterns in large SNNs is a critical research focus \cite{shainline2019superconducting, ceolini2020communication}. Integrating sensors directly (e.g., event cameras) also aids scalable real-time processing.

\textbf{Addressing System-Level Scaling Challenges:} Building brain-scale neuromorphic systems involves more than just packing components densely. System-level challenges such as efficient power delivery networks, effective thermal dissipation across large chip areas or multi-chip modules \cite{zhu2022comprehensive}, and robust fault tolerance mechanisms are crucial for reliable operation of systems with billions of components \cite{davies2021advancing}. Furthermore, efficient methodologies for mapping complex, large neural network architectures onto distributed neuromorphic hardware are essential for practical deployment \cite{roy2019towards}. Successfully tackling these system-level scaling issues is fundamental to moving neuromorphic computing from laboratory prototypes to powerful, real-world computational tools \cite{schuman2022opportunities}.

\subsection{Core Principles for Next-Generation SNN-Centric AGI}

\subsubsection{Synaptic Plasticity Mechanisms}
	At the heart of brain-inspired AGI is synaptic plasticity, the dynamic ability of neural connections to adapt and strengthen based on activity and experience. Modern spike-timing dependent plasticity (STDP) rules now integrate spike-timing over extended windows, further improving the stability of credit assignment in deeper SNNs \cite{gebhardt2024time}. Higher-order STDP formulations extend the classical pair-based rule by accounting for interactions among triplets of spikes and voltage-dependent mechanisms in synaptic updates to more faithfully capture cortical learning dynamics \cite{rodriguez2024enhancing}. Neuromodulation-inspired gates, modeled on dopaminergic reward signals, selectively reinforce synapses during salient events, enabling experience-driven consolidation \cite{rodriguez2024enhancing}. Meta-plasticity (learning to learn) frameworks dynamically adjust plasticity parameters during training, accelerating adaptation to new tasks and reducing catastrophic forgetting (the phenomenon where learning new tasks disrupts or overwrites previously stored synaptic patterns)\cite{rodriguez2024enhancing}. Finally, hardware-aware plasticity co-designs update rules with memristive device physics, enabling on-chip learning that is both energy efficient and resilient to device variability \cite{wang2025topology,putra2024hasnas}. 
\subsubsection{Efficient Neural Coding Strategies}
	Energy-efficient AGI requires sparse spike-based codes, where only a small fractions of neurons will fire on a given input. By leveraging heterogenous neuronal time-constraints, sparsity and robustness are improved in edge deployments \cite{chakraborty2024probabilistic}. Temporal codes, such as time-to-first-spike and rank-order schemes, encode features in the timing of a single spike, thereby slashing communication overhead \cite{pabian2024iteration}. Event-based sensor frontends (e.g., DVS cameras, spiking IMUs) generate asynchronous streams perfectly aligned with SNN processing and achieve microsecond latency and minimal redundancy \cite{bian2024evaluation}. Homeostatic normalization adapts synaptic strengths and firing thresholds to maintain stable acticity levels during continual learning and prevent pathological excursions \cite{chakraborty2024probabilistic}. Finally, information-theoretic design, which maximizes mutual information under strict energy budgets, guides the selection of coding schemes for optimal efficiency \cite{kausar2024novel}. 
\subsubsection{Multimodal Association and Dynamic Routing}
	True AGI will benefit from integrating diverse modalities. Spiking cross-modal transformers combine event-based audio and vision through tensor-fusion and attention blocks, demonstrating zero-shot audiovisual recognition \cite{li2024unified}. Contrastive objectives align representations across modalities, improving resilience to missing or noisy channels \cite{li2024unified}. Spiking mixture-of-experts models employ sparse gating to route inputs through specialized subnetworks and achieve conditional computation with minimal spikes \cite{xing2024spikellm}. In embodied settings, grounded language learning ties linguistic commands to sensorimotor trajectories and enables few-shot generalization in autonomous navigation \cite{liao2024integrating}. Contextual binding mechanisms, inspired by cortical microcircuits, dynamically allocate processing to salient features and support flexible reasoning and compositionality \cite{li2024organizational}. 
\subsubsection{Evaluation and Benchmarking Frameworks}
	Reproducibility hinges on standardized benchmarks. To compare SNNs against DNNs, key metrics such as accuracy, spikes-per-inference, and energy-per-inference are used \cite{xie2024toward}. Compression and pruning surveys analyze trade-offs between model size, inference latency, and spike efficiency, outlining the best practices for deployment on resource-limited platforms \cite{xie2024toward}. Common data-sets for event-based models include N-ImageNet, CIFAR-10-DVS, N-CARS, and SHD/SSC for classification under neuromorphic constraints \cite{bian2024evaluation}. The MMLU benchmark, typically used to test LLMs in various reasoning tasks, can also be applied to SNNs \cite{hendrycks2020measuring}. Similarly, NeuroBench evaluates SNNs on spanning sensory, motor-control, and reasoning tasks \cite{NeuroBench}, while SNNBench tests SNNs on both training and inference time on varied workloads \cite{BenchCouncilSNNBench2024}. 
    
    Furthermore, N‑MNIST and \textbf{N‑Caltech101} are spiking versions of the classic MNIST and Caltech101 image corpora created via sensor saccades \cite{orchard2015converting}. Finally, the community is beginning to converge on umbrella suites such as \textbf{NeuroBench}, a collaborative, MLPerf‑style framework that unifies tasks across vision, audio, continual learning, and control while standardising energy– and spike‑efficiency metrics \cite{yik2025neurobench}.

\subsubsection{Cross-cutting Considerations}
	Beyond core principles, AGI systems must ensure adversarial robustness, out-of-distribution generalization, and lifelong adaptability. Techniques such as adversarial STDP training and continual homeostatic updates decrease vulnerability to malicious or novel inputs \cite{xie2024osworld}. Hybrid learning pipelines unlock ultra-deep architectures with stable training dynamics \cite{gebhardt2024time}. Finally, ethical safeguards and transparency mechanisms must be embedded throughout to ensure trustworthy AGI in safety-critical domains \cite{rodriguez2024enhancing}.

\subsection{Brain-Computer Interfaces \& Cognitive Synergy}
Although brain-computer interfaces (BCIs) have made significant strides, they have yet to achieve the seamless, accurate, and non-invasive interaction between users and computers that is envisioned. This section delves into the current limitations, emerging research trends, the concept of cognitive synergy in BCIs, overarching goals, and the integration of artificial general intelligence (AGI) into BCI systems.
\subsubsection{Current Shortcomings in BCIs
}
Despite advancements, BCIs face several challenges:
\begin{itemize}

    \item Signal Quality and Reliability:
    Non-invasive BCIs often suffer from low signal-to-noise ratios, making it difficult to accurately interpret neural signals \cite{gouret2024advancements}.

    \item Invasiveness and Safety Concerns: 
    While invasive BCIs provide better signal fidelity, they pose risks such as infections, inflammation, and long-term biocompatibility issues \cite{kubben2024invasive}.

    \item User Comfort and Accessibility: 
    Many BCI systems are bulky or require extensive setup, limiting their practicality for daily use. \cite{drew2024united}

    \item Ethical and Privacy Issues:
    The potential for unauthorized access to neural data raises significant ethical concerns regarding user autonomy and data security \cite{kubben2024invasive}.
\end{itemize}

\subsubsection{Current Trends in BCI Research}
Recent research in BCIs focuses on several key areas:
\begin{itemize}
    \item Minimally Invasive Technologies: Companies like Synchron have developed stent-like devices that can be implanted via blood vessels, reducing surgical risks \cite{gouret2024advancements}.
    \item Integration with AI and Machine Learning: Advanced algorithms are being employed to enhance signal decoding and adapt to individual neural patterns, improving BCI performance \cite{drew2024united}.
    \item Multi-Modal Interfaces: Combining BCIs with other input methods, such as eye tracking or muscle signals, to create more robust and versatile systems \cite{gordon2024ethical}.
    \item Clinical Applications: BCIs are increasingly being used in therapeutic settings, aiding in neurorehabilitation and providing communication means for individuals with severe motor impairments \cite{eze2024applications}.
\end{itemize}
\subsubsection{Concept of Cognitive Synergy in BCIs}
Cognitive synergy refers to the enhanced cognitive capabilities achieved through the integration of human intelligence with machine processing. In the context of BCIs, this involves creating systems where human neural inputs and artificial intelligence work in tandem to perform tasks more comprehensively than either could alone. This framework will synergize human decision-making, learning, and problem-solving abilities with machine intelligence \cite{george2025enhancing}.

\subsubsection{AGI for BCIs}
The integration of AGI into BCIs holds the promise of creating systems capable of understanding and adapting to a wide range of tasks and environments. AGI-enhanced BCIs could facilitate more natural interactions, allowing for real-time language translation, complex problem-solving, and adaptive learning. This fusion aims to create a symbiotic relationship where human intuition and machine intelligence complement each other, leading to unprecedented levels of efficiency and innovation.
\subsubsection{Overall Goals for BCIs
}
Some of overall objectives driving BCI development include:
\begin{itemize}
    \item \textbf{Restoration of Function:} Enabling individuals with neurological disorders or injuries to regain lost motor or sensory functions \cite{ruiz2024emerging}.
    \item \textbf{Enhancement of Human Capabilities:} Augmenting cognitive or physical abilities beyond natural limits \cite{pan2024towards}.
    \item \textbf{Seamless Human-Computer Interaction:} Creating intuitive interfaces that allow for effortless communication between humans and machines \cite{mukhtar2025artificial}.
    \item \textbf{Neurotherapeutic Applications:} Utilizing BCIs for diagnosing and treating mental health conditions, such as depression or anxiety \cite{khorev2024review}.
\end{itemize}
\subsection{Ethics \& Trustworthiness}
Large language models (LLMs) are increasingly employed in everyday life, raising issues regarding their ethical use and trustworthiness. They're already appearing in critical fields such as education, healthcare, and law, where even little errors or skewed conclusions can have major effects.

The TrustLLM benchmark was developed to assess the trustworthiness of LLMs \cite{huang2024position}. It evaluates models on six dimensions: honesty, safety, fairness, robustness, privacy, and machine ethics. These regions mirror real-world values, which influence how users interact with and rely on AI systems. The benchmark assessed 16 LLMs across over 30 datasets. While proprietary models such as GPT-4 and Claude 2 performed well overall, open-source models such as LLaMA-2 and ChatGLM excelled in specific areas. Interestingly, the study discovered that being more transparent does not always indicate a model will perform poorly \cite{huang2024position}.

A noteworthy finding was that some models classified safe reactions as harmful. While this demonstrates caution, it may impede usage in sensitive areas such as mental health support or content moderation. Another consideration is justice. Certain models still exhibit demographic bias or provide unequal responses based on race or gender, raising ethical concerns \cite{huang2024position}. 

Furthermore, trust is also determined by the robustness of models in the face of adversarial stimuli or misinformation. Many models remain susceptible to these inputs. Some were easily manipulated into generating false or objectionable results. This demonstrates that further technical advancements are required before LLMs can be reliably used at scale.

If these issues aren't resolved, people may lose interest in AI. Developers and researchers are responsible for keeping systems in check and making adjustments. Ethics and trust should not be optional; rather, they are critical to how LLMs affect individuals and society. Benchmarks such as TrustLLM help to keep models responsible and guide AI progress.

\subsection{Open Challenges}
	As SNNs and neuromorphic hardware scale toward general-purpose intelligence, several interdisciplinary hurdles remain at the intersection of neuroscience, deep learning, and hardware engineering. We highlight five pressing challenges.

\subsubsection{Spiking-First Foundation Architectures}
Integrating spiking-neuron dynamics, such as temporal coding and asynchronous event handling, into transformer-scale models remains a central challenge, since modern models rely on high-precision arithmetic and dense activations. Early efforts like SPikeLLM adapt saliency-based spiking attention to billion-parameter language models, achieving competitive perplexity with 10x fewer operations \cite{Xing2024,Zhu2023}. However, the full integration of spiking neuron dynamics (e.g., temporal coding and asynchronous event handling) into transformer-scale architectures while maintaining LLM capabilities remains unsolved. 

As predicted, a robust step toward this goal is the design of robust mixed-signal fabrics that couple analog neuromorphic cores (which excel at event processing) with digital control logic (which enables programmability). Intel's Loihi pioneered this coupling, but integrating tight feedback loops between analog neuron states and digital learning engines while mitigating device mismatch and noise remains a challenge \cite{Davies2018,pedersenNeuromorphicIntermediateRepresentation2024}. Scalable hardware in-the-loop engines for surrogate-gradient updates are needed. 

Looking further ahead, emerging substrates, such as photonic and spintronic hardware, promise GHz-rate spiking with minimal energy consumption. Early photonic SNN demos, for instance using phase-change synapses, showcased fully optical integration and fast signal processing \cite{Feldmann2019,Cheng2024}. Building photonic/spintronic spiking transformer prototypes with scalable memory, on-chip STDP or surrogate‐gradient updates, and reliable integration remains an open hardware frontier.

\subsubsection{Scalable Learning and Credit Assignment}
	Assigning credit across long temporal dependencies is straightforward in DNNs via backpropagation through time (BPTT), but BPTT is neither scalable nor biologically plausible in large SNNs. Surrogate-gradient BPTT accelerators and forward-mode differentiation methods (e.g., e-prop, FPTT) perform well on small benchmarks but degrade on tasks requiring thousands of timesteps \cite{Yin2023,Cramer2022}. Efficient, low-memory algorithms for long-range spike-based credit assignment, potentially inspired by three-factor plasticity, are needed. 
     
     While gradient‐based SNN training thrives in software, embedding full surrogate-gradient or temporal-difference learning engines on neuromorphic chips remains a challenge. Proof-of-concept implementations of on-chip backprop (Loihi) and e-prop (SpiNNaker 2) demonstrate feasibility \cite{Renner2024,Rostami2022}, but scaling to multi‐layer deep networks with billions of parameters in real time demands new local circuits for error propagation and spike-time TD updates.

\subsubsection{Hierarchical Memory and Event Routing}
	Brains organize memory hierarchically, from synaptic eligibility traces to hippocampal indices to cortical consolidation, but SNNs lack analogous modules. Spiking transformer variants like the SpikingResformer and One-Step Spiking Transformer begin to introduce multi‐layer attention in a few timesteps \cite{shi2024spikingresformer,ijcai2024p348}, but co-designing explicit spike-based short- and long-term memory arrays and integrating them into attention architectures remains a challenge.
     
     Although attention excels at routing information across layers, it has not been co-designed with explicit spike-memory modules. MoE‐style spiking architectures (e.g., Spiking Transformer with Experts Mixture) demonstrate sparse routing \cite{Zhou2024}, but end-to-end designs integrating multiple spike-memory caches (e.g., replay buffers, synaptic caches) within an attention stack are still missing.

      At the hardware level, scaling SNNs requires reliable, low-latency spike routing across chips. Existing NoC architectures address congestion but lack guarantees on delivery order and jitter required for real-time AI tasks \cite{ostrauBenchmarkingNeuromorphicHardware2022,Koesters2023}. Standardized cross-chip protocols that preserve spike timing fidelity and support event compression are needed.

 Finally, implementing transformer‐style attention purely with spikes, without typical global clocks or floating‐point operations, requires novel sparse routing algorithms. Saliency‐based gating in SpikeLLM offers one blueprint \cite{Xing2024}. However, real‐time, event‐driven attention at a billion-parameter scale will demand further innovations such as ultra-low-latency spike dispatch, dynamic load balancing, and fault tolerance across cores.

 \subsubsection{Lifelong and Embodied Intelligence}
  Local plasticity rules mitigate forgetting in small-scale tasks \cite{Minhas2024}, yet SNNs still struggle to learn long sequences of tasks at scale. Synaptic consolidation, dynamic resource allocation, and meta-plasticity mechanisms must be co-developed to sustain high accuracy across dozens of tasks without manual intervention.

   Embodied cognition demands a tight coupling of linguistic processing with sensorimotor control. Spiking RL controllers manage simple motor tasks \cite{Zanatta2024}, and spiking LLMs (BrainGPT) handle text \cite{Tang2024b}, but no unified framework yet encodes both modalities in a common spiking format. Designing hybrid AGI architectures that operate seamlessly on text and real-world sensory streams is a challenge.

\subsubsection{Ethical Safeguards}
 As neuromorphic AI enters critical domains (self-driving cars, medical devices), embedding ethical guardrails at the hardware/software boundary is essential. While high-level AI ethics frameworks exist \cite{Floridi2023}, translating policies into on-chip monitors, safe-plasticity limits, and transparent spiking decision logs remains unexplored. Hardware-enforced “kill switches” or bias-detection circuits may be required to ensure trust in autonomous neuromorphic agents.

\subsection{Educational Perspective}
This work was mainly undertaken as a collaborative project by students in our Computational Neuroscience course at The University of Georgia, illustrating a pedagogical approach that transcends traditional knowledge transmission to enhance the cultivation of critical thinking, innovation, and teamwork. By engaging directly with complex interdisciplinary research—spanning neurobiological theory, algorithmic design, and neuromorphic hardware implementation—students develop the ability to decompose grand challenges into tractable subproblems, evaluate competing hypotheses, and select appropriate experimental or simulation frameworks. The integrative perspective advanced in this paper offers a compelling model for future education, synergizing the convergence of neuroscience, artificial intelligence and hardware design as a unified learning domain. By studying how synaptic plasticity, spike-based communication and multimodal association inform the co-design of algorithms and neuromorphic substrates, students develop a deep, research-driven understanding of both theoretical principles and practical implementations. Embedding these interdisciplinary insights into the curriculum encourages learners to formulate biologically inspired hypotheses, translate them into computational experiments, and prototype on emerging hardware platforms, all within a collaborative framework that mirrors real-world research teams. Regular cycles of peer critique, faculty mentorship and academic/industry collaborations reinforce scientific rigor, ethical reflection and clear technical communication. In turn, this pedagogy fosters adaptive expertise, systems thinking and innovative problem-solving skills, equipping students and their collaborators/mentors to navigate and shape the rapidly evolving landscape at the intersection of brains and machines. By continuously aligning coursework with the frontiers delineated in this paper, educational programs can cultivate lifelong learners capable of driving the next wave of breakthroughs.

\section{Conclusion}\label{Section 9}

\subsection{Integrative Agenda Summary}
In this survey we have traced parallel trajectories in neuroscience, artificial general intelligence via large language models, and neuromorphic computing, and shown how these intersections integrated to a unified research agenda. This agenda rests on four pillars:
\begin{enumerate}
    \item \textbf{Co-Design of Brains, Algorithms and Hardware:} Neuroscience delivers the mechanistic motifs that machine-learning researchers can translate into scalable learning objectives and data-efficient inductive biases. Meanwhile, neuromorphic platforms offer silicon substrates through which these motifs can run at brain-like energy efficiency in real time. 
    \item \textbf{Hybrid Learning Pipelines:} The most promising systems combine DNN optimizations for pursuing the global optimum with spike-based fine-tuning or neuromodulated adaptation for continuous learning and lifelong plasticity. Such hybrids are already powering large language models (SpikeLLM, SpikeGPT) that reach DNN-level perplexity at orders of magnitude lower energy cost.
    \item \textbf{Hierarchical Memory and Sensorimotor Grounding:} The combination of hippocampal-inspired replay buffers, attractor memory in cortical SNN modules, and attention-gated long-context windows suggests a multi-timescale hierarchy of memory that is capable of episodic recall and semantic abstraction. Embodied agents that combine event-driven sensorimotor cortices with spiking language models point toward a fully grounded AGI. 
    \item \textbf{Standardization and Benchmarking:} The emergence of NeuroBench, ONNX-SNN converters, and cross-chip spike-routing protocols give the community common metrics for accuracy, latency, spikes-per-inference, and energy consumption. These benchmarks and tools are critical both for comparing research results fairly and for moving designs from the lab into real-world applications.
\end{enumerate}

\subsection{Looking Forward \& Final Thoughts}
As we look ahead, several themes will shape the coming years:
\begin{itemize}
    \item \textbf{From components to systems:} We must move beyond isolated demonstrations (e.g., spiking transformers, on-chip STDP) toward fully integrated neuromorphic AGI prototypes that combine learning memory, perception, and action into a single platform.
    \item \textbf{Closing the loop with biology:} Advances in connectomics, brain organoids, and large-scale brain simulations will provide rich blueprints for computation. In turn, neuromorphic AGI systems will become indispensable tools to test these neuroscientific hypotheses at scale and in real time. 
    \item \textbf{Responsible and ethical deployment:} As neuromorphic AGI moves into the fields of robotics, healthcare, and edge intelligence, we must embed ethical safeguards such as transparency and robust fail-safes at the hardware level. 
    \item \textbf{Democratizing neuromorphic tools:} To realize this vision broadly, a community-driven framework of open software stacks, standardized benchmarking, and accessible hardware testbeds will accelerate innovation and lower the barriers for interdisciplinary teams. 
\end{itemize}

The pursuit of human-level intelligence requires a seamless integration of neuroscience, computational algorithms, and hardware design. By embracing an integrative co-design approach, we can develop adaptive, energy-efficient systems that not only advance artificial intelligence but also deepen our understanding of the human mind.


\bibliographystyle{unsrt}
\bibliography{main_bib, sample-base}


\end{document}